\documentclass[iop]{emulateapj}
\usepackage{mathrsfs}
\usepackage{color}
\usepackage{amsmath}
\usepackage{multirow}
\usepackage{hyperref}
\usepackage[normalem]{ulem}

\def\bd{\begin{displaymath}} \def\ed{\end{displaymath}}
\def\ba{\begin{aligned}} \def\ea{\end{aligned}}
\def\nms{\mathsurround=0pt} \def\oversim#1#2{\lower
4pt\vbox{\baselineskip 0pt \lineskip 1pt
\ialign{$\nms#1\hfil##\hfil$\crcr#2\crcr\sim\crcr}}}
\def\ga{\mathrel{\mathpalette\oversim>}}
\def\la{\mathrel{\mathpalette\oversim<}} 
  \def\obs{{\rm obs}}
\def\bh{{M_{\bullet}}} \def\opt{{\rm opt}} 
  
\def\msun{M_{\odot}}  
  \def\kms{\rm ~km~s^{-1}}

\def\be{\begin{equation}} \def\ee{\end{equation}}

\begin{document}
\title{Extracting the Possible Intrinsic Relation between Radiative Efficiency and Mass of QSOs: a Maximum Likelihood Method and its Application to the SDSS DR7 QSOs}
\author{Fupeng Zhang$^{1,2,3,4}$ and Youjun Lu$^{5,6}$}
%\tableofcontents
\affil{ $^1$\,School of Physics and Materials Science, Guangzhou
University, 510006 Guangzhou, China, zhangfupeng@gzhu.edu.cn\\
$^2$\,School of Physics and Astronomy, Sun Yat-Sen University,
Guangzhou 510275, China\\
$^3$\,Key Laboratory for Astronomical Observation and Technology of Guangzhou, 510006 Guangzhou, China\\
$^4$\,Astronomy Science and Technology Research Laboratory of Department of Education of Guangdong Province, Guangzhou 510006, China\\
$^5$\,CAS Key Laboratory for Computational Astrophysics, National Astronomical Observatories, Chinese Academy of
Sciences, Beijing, 100101, China; luyj@nao.cas.cn  \\
$^6$\,School of Astronomy and Space Sciences, University of Chinese
Academy of Sciences, No. 19A Yuquan Road, Beijing 100049, China
}

\begin{abstract}
Radiative efficiencies of QSOs and its distribution encode rich information on the evolution of both masses and spins of massive black holes (MBHs) across cosmic time. In this paper, we develop a maximum likelihood method to statistically extract the intrinsic relation between radiative efficiency ($\epsilon$) and mass ($M_{\bullet}$) of QSOs from their distribution on the luminosity-(empirically estimated virial) mass plane. By using mock samples, we find that strong constraint can be put on the $\epsilon-M_{\bullet}$ relation at redshift $z\lesssim 0.4$ from uniform QSO samples similar to those in Sloan Digital Sky Survey, and from QSO samples at $z \sim 0.6$  (or $\lesssim 1.0$) if the magnitude limit of the survey can  be $\sim 1-2$ (or $2-3$) magnitude deeper. Applying this method to the SDSS DR7 QSOs with $z\la 0.7$, we find $\epsilon \propto M_{\bullet}^{0\sim 1.1}$ (or $\epsilon \propto M_{\bullet}^{-1.0\sim 0}$) correlation for QSOs with the masses obtained according to the H\,$\beta$ (or Mg\,II) empirical mass estimator. These contradictory results may be due to the unknown systematic errors in the two mass estimators, preventing an accurate constraint on the $\epsilon-M_{\bullet}$ relation by using current available QSO samples. We find that both the estimates of MBH mass and Eddington ratio distribution functions can be affected by the $\epsilon-\bh$ relation, suggesting that the determination of this relation is important for understanding the accretion and growth history of MBHs. In future, the intrinsic $\epsilon-M_{\bullet}$ relation is expected to be strongly constrained by using QSO samples obtained from surveys deeper than SDSS if the host galaxy contamination and systematic errors of the mass estimator(s) can be well modeled or removed.
\end{abstract}


\section{Introduction}
There is a consensus that massive black holes (MBHs; in the mass range $M_{\bullet} \sim10^5-10^{10}\msun$) lie in the centers of most galaxies. The mass growth and spin evolution of these MBHs are mainly governed by their accretion histories \citep[e.g.][]{Soltan82, YT02, Marconi04, YL04, kin06, per09, Volonteri13, ZL19}. Most QSOs are probably accreting gaseous material via the standard thin disk \citep{Shakura73, NT73}, and the mass growth rate of their central MBHs and the energy radiated from them are determined by the accretion rate and radiative efficiency, with the latter one directly determined by the MBH spin. MBHs with different masses are likely residing in different types of galaxies and environments, for example, MBHs at the high-mass end mostly reside in giant ellipticals, while those at the low-mass end mostly reside in spiral bulges, and thus they may have different accretion histories. A correlation between radiative efficiency $\epsilon$ (or spin $a$) and mass of active MBHs may emerge because of the environmental differences \citep[e.g., coherent or chaotic accretion][]{Dotti13, Dubois14, ZL19}. Such a correlation, if exists, may be a probe to the accretion histories of MBHs.

Currently, it is still not clear whether there is an intrinsic $\epsilon-M_{\bullet}$ correlation or not. \citet{Davis11} directly estimated the radiative efficiencies for PG QSOs by $\epsilon = L_{\rm bol}/\dot{M}_{\rm acc} c^2$, where $L_{\rm bol}$ is the bolometric luminosity,
$\dot{M}_{\rm acc}$ the rate inferred from the standard thin accretion disk modeling, and $c$ the speed of light. They found a correlation between the radiative efficiencies and MBH masses of those PG QSOs. \citet{Wu13} further estimated the radiative efficiencies for SDSS QSOs and also found such a correlation, but they attributed it to an apparent correlation that is possibly induced by the sample flux limit and uncertainties in the MBH estimates \citep[see also][]{Raimundo12}. Therefore, it is important to model and remove those effects induced by the sample selection and measurement errors in the $\epsilon-M_\bullet$ relation analysis, in order to confirm whether there is an intrinsic correlation or not.

In the standard thin accretion disk model, the luminosity of a QSO at the optical-band depends on the accretion rate, MBH mass, as well as spin.
For systems with a fixed Eddington ratio $\lambda$ (e.g., $0.01-1$), the mass accretion rate decreases with increasing radiative efficiency, and the optical-band luminosity is correspondingly affected.
To estimate the optical-band luminosities accurately, the relativistic correction is important if the spin $a$ (and correspondingly $\epsilon$) is large, and it is even substantial if $M_{\bullet} >10^9M_\odot$ because the peak of spectral energy distribution (SED) shifts to the optical-band and a larger $\epsilon$ corresponds to a higher peak.
For these reasons, if
there is an underlying intrinsic  $\epsilon-M_{\bullet}$ relation for a large sample of QSOs,
their distribution on the luminosity-MBH (virial) mass plane may be significantly different from that of a sample without such an intrinsic relation.

In this paper, we develop a maximum likelihood method to extract the intrinsic $\epsilon-M_{\bullet}$ relation, if any, from the distribution of a large number of QSOs on the luminosity-MBH (virial) mass plane, by utilizing the relativistic thin accretion disk model \citep[e.g.,][]{NT73}.
We take into account both the selection effects and the uncertainties in the MBH mass estimates. We first test the validity of this method by using mock samples of QSOs with different settings on the magnitude/flux limit, and then apply it to the SDSS DR7 QSO sample to obtain constraint on the possible intrinsic $\epsilon-M_{\bullet}$ relation.

This paper is organized as follows. In Section~\ref{sec:analysis}, we develop an analytical approach to calculate the optical band luminosity of individual QSOs by adopting the relativistic standard thin accretion disk model. Assuming such a model, we then introduce a maximum likelihood method in Section~\ref{sec:maxlkhood} for extracting the possible intrinsic $\epsilon-M_{\bullet}$ relation. We generate a number of mock QSO samples and use the Markov Chain Monte-Carlo (MCMC) fitting technique to verify this method in Section~\ref{sec:mock_test}. In Section~\ref{sec:applying_to_sdss}, we apply this method to the SDSS DR7 QSO sample and obtain constraints on the $\epsilon-M_\bullet$ relation, MBH mass function (BHMF), and Eddington ratio distribution function (ERDF). Discussions and conclusions are given in Section \ref{sec:dis} and \ref{sec:con}, respectively.

In this paper, we assume a flat $\Lambda$CDM cosmology with parameters ($h_0, \Omega_{\rm m}, \Omega_\Lambda)=(0.7, 0.3, 0.7)$, where $h_0=H_0/100$\,km s$^{-1}$\,Mpc$^{-3}$ with $H_0$ as the Hubble constant, $\Omega_{\rm m}$ and $\Omega_{\Lambda}$ are the fractions of matter and cosmological constant in the local universe, respectively.

\section{Relativistic thin accretion disk model: an analytical approach}
\label{sec:analysis}

\begin{figure}
\center
\includegraphics[scale=0.9]{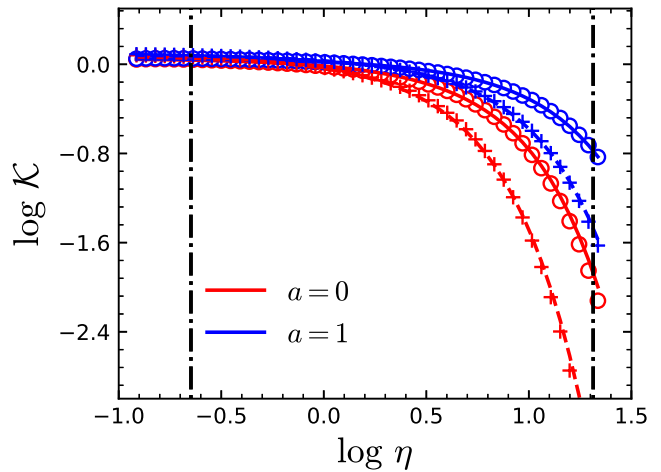}
\caption{The correction factor $\mathcal{K}$ as a function of $\eta$ (see definition of $\mathcal{K}$ in Eq.~\ref{eq:loptad}). Open circles and plus symbols show $\mathcal{K}$ that  obtained by numerically integrating Equation~\eqref{eq:lnu}, and lines show the the fitting results given by Equation~\eqref{eq:kappa}. The solid and dashed lines show the results obtained by assuming  $f_{\rm col}=1.7$ and $1$, respectively.
The red and blue lines show the results obtained for $(\bh, \lambda, \epsilon)= (10^6\msun, 1, 0.057)$ and $(10^{11}\msun, 0.01,0.4)$, respectively. The values of $\eta$ for bright QSOs should be within the region confined by these two vertical dot-dashed lines.
}
\label{fig:kappa}
\end{figure}

In this section, we introduce an analytical model to estimate the optical luminosity radiated from a system that accretes material via the standard thin disk. This analytical method is calibrated by fitting to the numerical results obtained from the sophisticated relativistic accretion disk model implemented with the ray-tracing technique. Details for the ray-tracing method are provided in a previous work \citep[see][]{Zhang15}. By adopting such a numerical method, we trace back each ray in a distant observer's image plane to the accretion disk in the equatorial plane of the MBH. Below we show only the results obtained from the accretion disk models at the wavelength $2500$\AA\, as it will be applied to the SDSS-like QSO surveys.

The optical luminosity of a QSO in the standard thin accretion disk model depends on the MBH mass $\bh$, accretion rate ($\dot{M}_{\rm acc}$; or Eddington ratio $\lambda$), efficiency ($\epsilon$, or the MBH spin $a$), and the inclination angle ($i$, defined as the angle between disk normal direction and line of sight). We adopt the ray-tracing method to obtain the observed luminosity at any frequency ($\nu$) by integrating over the surface of accretion disk
\be
L_{\nu}=4\pi \int g^3 I_{E0}(r_{\rm e},\theta_{\rm e})d\alpha d\beta.
\label{eq:lnu}
\ee
Here $\alpha$ and $\beta$ are the two impact parameters describing the position of a disk element at ($r_{\rm e}$, $\theta_{\rm e})$ on the observer's sky.
$g=E/E_{\rm e}$ is the relativistic correction factor~\citep[For more details see][]{Cunningham75}, where $E_{\rm e}$ is the energy of a photon at the rest frame of the disk element and $E$ is the energy of the photon received by the observer. $I_{E_0}$ is the intensity at the rest frame of the disk element that emits the photon, which is given by
\be
I_{E_0} = f_{\rm col}^{-4}F(\nu).
%
%\ea
%
\ee
Here $f_{\rm col}$ is the spectral hardening factor \citep{Shimura95,Salvesen13}, $F(\nu)$ the intensity of a black body radiation at temperature $T$, where
\be
T = f_{\rm col}\left(\frac{3c^6}{8\pi G^2\sigma}\right)\frac{{\dot{M}_{\rm acc}}^{1/4}}{\bh^{1/2}}Q^{1/4}(x),
%
%\ea
%
\ee
with $G$ the gravitational constant, $\sigma$ the Thompson scattering cross section.
The function $Q(x)$, with $x=r_{\rm e}^{1/2}$ (in unit of $r_{\rm g} = GM_\bullet/c^2$), can be found in \citet{Gierlinski01}.

For X-ray binaries, the typical value for the hardening factor $f_{\rm col}$ is
$\sim 1.7$ \citep{Shimura95}. For QSOs, however, there are large uncertainties in the estimates of this parameter.
For simplicity, we only consider the following two cases: (1) $f_{\rm col}=1$, corresponding to the most simple standard thin accretion disk model \citep{NT73}, and (2) $f_{\rm col}=1.7$, considering the effect of Comptonization \citep{Shimura95}. We find that the resulting luminosity from the latter case is almost the same as that predicted by the TLUSTY model \citep{Hubeny95} for a system with the same physical parameters (the differences in predicted luminosities usually $<0.04-0.2$ dex).

The maximum likelihood method mentioned in Section~\ref{sec:maxlkhood} requires calculations of the (optical-band) luminosities of a large number of QSOs under various parameter settings. For a practical point of view, it may be more efficient if a fast and accurate analytical approximation to Equation~(\ref{eq:lnu}) can be obtained for such an implementation.

For most QSOs, the optical-UV band luminosities given by Equation~\eqref{eq:lnu} follow a simple scaling relation, i.e., $L_\nu \propto \lambda^{2/3} \epsilon^{-2/3} \bh^{4/3}  \cos i$,  when $h \nu \ll kT$ and the relativistic effects are ignored.  Therefore, we introduce an analytical form to approximate the optical band luminosity for a wide range of conditions, i.e.,
\be
\ba
L_{\nu}=c_0 f_{\rm col}^{-4/3}\mathcal{K}\lambda^{2/3}
\epsilon_{0.1}^{-2/3} M_{\bullet,8}^{4/3} \cos i.
\label{eq:loptad}
\ea
\ee
Here $\mathcal{K}$ is a correction factor by considering the relativistic effects and the bending of the black body radiation spectrum,
$\epsilon_{0.1} = \epsilon/0.1$ and $M_{\bullet,8}=\bh/10^8\msun$, $c_0$ is a constant and set to $c_0 = 2.54\times10^{45}$\,erg\,s$^{-1}$ in order to make $\mathcal{K}=1$ when $M_{\bullet,8}=\lambda=\epsilon_{0.1} =f_{\rm col}=\cos i=1$.

Using the analytical Equation~(\ref{eq:loptad}) to fit the numerical results obtained from Equation~(\ref{eq:lnu}) at wavelength $2500$\AA, we find that $\mathcal{K}$ can be approximated described by
\be
\ba
\mathcal{K}&\simeq\frac{u\eta}{\exp(w\eta)-1},
\label{eq:kappa}
\ea
\ee
where $\eta=M_{\bullet,8}^{1/2}/{\dot{M}_{\rm acc,2.2}}^{1/4} = \lambda^{-1/4}M_{\bullet,8}^{1/4} \epsilon_{0.1}^{-1/4}$, and $\dot{M}_{\rm acc,2.2} =\dot{M}_{\rm acc}/(2.2\msun / {\rm yr})$. The value of $u$ and $w$ depend only on the spin parameter and $f_{\rm col}$.  If assuming $f_{\rm col}=1.7$, we have
\be
\ba
u&=0.0355r_{\rm ISCO}/r_{\rm g}+0.131,\\
w&=0.0315r_{\rm ISCO}/r_{\rm g}+0.116.
\label{eq:kappa1}
\ea
\ee
If assuming $f_{\rm col}=1$, we have
\be
\ba
u&=0.0655r_{\rm ISCO}/r_{\rm g}+0.242,\\
w&=0.0553r_{\rm ISCO}/r_{\rm g}+0.196.
\label{eq:kappa2}
\ea
\ee
Here $r_{\rm ISCO}$ is the radius of the innermost stable circular orbit (ISCO)~\citep{Bardeen72}.

Figure~\ref{fig:kappa} shows both the numerical results of $\mathcal{K}$ as a function of $\eta$ and that given by the fitting formula (Eq.~\ref{eq:kappa}). The fitting formula well matches the numerical results with an error $\ll 0.05$\,dex.
In most conditions, $\eta\simeq 1$, and thus $\mathcal{K}\simeq1$. However, for QSOs with $\eta \gg 1$, i.e., if the MBH mass is large and the Eddington ratio is small, e.g., $\bh\sim10^{10}\msun$ and $\lambda \sim 0.01$, the correction factor $\mathcal{K}$ can be much smaller, i.e., $\lesssim 0.01$. $\mathcal{K}$ is also sensitive to the values of spin and $f_{\rm col}$ if $\eta\gg 1$.

According to Equation~(\ref{eq:loptad}), we have $L_\nu\propto \mathcal{K} \epsilon^{-2/3}$. Thus, if the correction factor $\mathcal{K}$ is close to unity, for any given $\lambda$, and $M_\bullet$, if varying $\epsilon$ from $0.4$ to $\sim 0.04$, the optical-UV luminosity $L_\nu$ can be reduced by $\sim 0.67$\,dex. For QSOs with MBHs at the high-mass end or small Eddington ratio, $\mathcal{K}$ becomes also significant. Therefore, the optical-UV luminosity depends significantly on the radiative efficiencies.
For individual QSOs, there is a degeneracy between the unknown $\lambda$ and $\epsilon$, as suggested by
Equation~\eqref{eq:loptad}, which prevents the determination of $\epsilon$ simply from observations on $L_\nu$ and $M_{\bullet,8}$.
%}
However, it may be done for a large sample of QSOs, as the joint distribution of the MBH masses and optical-UV-band luminosities depends on the underlying intrinsic correlations between MBH mass $M_{\bullet}$ and radiative efficiency $\epsilon$. The details of our method is shown in the following sections.

\section{A maximum likelihood method}
\label{sec:maxlkhood}

In this section, we introduce a maximum likelihood method that can be used to extract the possible intrinsic $\epsilon-\bh$ relation. We start with a general formalism of the method in Section\,\ref{subsec:likelihood}, and then consider more specific selection function(s) and probability distribution function (PDFs) of various QSO properties in Section\,\ref{subsec:select} and \ref{subsec:distribution}, respectively.

\subsection{The likelihood function}
\label{subsec:likelihood}

We denote the joint PDF of QSOs as $\Phi(\bh, L_{\nu}, z)$, which describes the number density of QSOs with luminosity in the range from $L_\nu$ to $L_\nu +dL_\nu$, true MBH mass in the range from $\bh$ to $\bh + d\bh$, and redshift in the range from $z$ to $z+dz$.
For most QSOs, $L_\nu$ and $z$ can be directly measured with high accuracy. However, the observational determined MBH mass $M_{\bullet,\rm obs}$ may substantially deviate from the true MBH mass $\bh$ and this deviation may be denoted by a PDF $P(M_{\rm obs}|\bh)$. Therefore, the observational determined joint PDF of QSOs, $\Phi_{\rm o}(M_{ \bullet,\rm obs}, L_{\nu}, z)$, is given by
\be
\Phi_{\rm o}(M_{\bullet,\rm obs}, L_{\nu},z) = \int \Phi(\bh,
L_\nu,z)P(M_{\obs}|\bh) d\bh.
\label{eq:phiomobs}
\ee
Assuming that the selection of QSOs depends only on the flux/luminosity limit of a QSO survey, and the selection function can then be denoted by $\Omega(L_{\nu}, z)$. Note that this assumption may be a good approximation for the SDSS QSO survey, however, it may be an oversimplification for QSO surveys in general. In order to extract the intrinsic joint PDF, we define a likelihood function $\mathcal{L}$ as \citep[see][]{Marshall83}\footnote{In some cases, $\Phi_{\rm o}=0$ as the input parameters of the model poorly describe the observed (or mock) samples. For example, the scatter of the observational determined MBH masses around the true MBH masses $\sigma_{M_\bullet}$ given by Eq.~(\ref{eq:pselect}) below is set too small to explain the observed mass distribution of QSOs. For these cases we set $\Phi_{\rm o}$ to be a tiny value to avoid odd results.}
\be
\ba
\ln \mathcal{L} &= \sum_{i=1}^N\ln \Phi_{\rm o}(M_{{\rm obs},i},
L_{{\nu},i},z_i)\\ &-\iiint \Phi(\bh, L_{\nu},
z)\Omega(L_{\nu},z)\frac{dV}{dz}dzd\bh dL_{\nu},
\label{eq:maxim_likelihood}
\ea
\ee
where $N$ is the total number of QSOs in the observational sample. The first term in the right side of Equation~\eqref{eq:maxim_likelihood} sums over all the observed QSOs giving their individual masses $M_{\bullet,{\rm obs},i}$, luminosity $L_{\nu,i}$, and redshift $z_i$, and the second term integrates over $\bh$, $L_\nu$, and $z$. It is practically convenient to use the joint distribution of MBH and Eddington ratio, i.e., $\Phi(\bh,\lambda,z)$ rather than $\Phi(\bh,L_{\nu},z)$ in modeling the intrinsic distribution of QSOs. According to Equation~\eqref{eq:loptad}, the joint PDF $\Phi(\bh,L_{\nu},z)$ can be also described as
\be
\Phi(\bh, L_{\nu},z)
=\frac{\Phi(\bh,\lambda,z)}{\left|\frac{\partial L_{\nu}}{\partial \lambda}\right|}.
\ee
We can then calculate Equation~\eqref{eq:maxim_likelihood} according to the assumed joint distribution of $\Phi(\bh,\lambda,z)$ and the intrinsic $\epsilon-\bh$ relation.

\subsection{Observational bias and selection functions}
\label{subsec:select}

For a given emission line, assuming that the true mass of the MBH $\bh$ can be estimated by an ideal empirical relation\footnote{We assume that the mass estimator is a function of $L_\nu$, which is used for the selection in Equation~\eqref{eq:maxim_likelihood}. Usually, the mass estimator is a function of the continuum luminosity around the emission line $L_{\rm em}$, which may be different from $L_\nu$ (at $2500\AA$). Considering that the continuum spectrum of a QSO can be  described by a power law, $L_\nu$ can always be obtained according to a given $L_{\rm em}$, and the difference between these two luminosities is then a constant that can be absorbed into the parameter $\bar b$.
},  e.g.,
\be
\ba
\log\bh&=\bar a+\bar b\log L_\nu +\bar c \log{\rm\,FWHM},
\ea
\ee
where $\bar a$, $\bar b$ and $\bar c$ are constants, FWHM is the full width at the half maximum of the adopted emission line. 
Due to various observational uncertainties, the empirical mass estimator may be not accurately obtained, so that it is 
scattered around, or biased from the true ones \citep[e.g., due to stochastic dispersions in luminosity, see][]{Shen13}. As a result, the observed mass $M_{\bullet,\rm obs}$ of QSOs given by the empirical mass estimator for QSOs now becomes
\be
\log M_{\bullet,\rm obs}=a+b \log L_\nu +  c \log{\rm\,FWHM} + s, 
\ee
where $s$ is a random deviation following a Gaussian distribution with a zero mean and a scatter of $\sigma_{M_\bullet}$.

In this paper, the adopted mass estimator is based on either the H$\beta$ or MgII line, 
thus, following \citet{Shen11} we set $c=\bar c=2$, assuming that both of these two emission lines are coming from the same regions of the 
broad line region (BLR; see~\citet{Onken08} for details.). Under such assumption, we have ignored the complexities of any systematic errors dependent on FWHM. 
\footnote{However, we notice that observations do suggest that $\bar c$ may be biased
and not exactly $2$ \citep[e.g.,][]{Wang09}. For the possible systematic errors of FWHM and 
its impact on our results, see discussions in Section~\ref{subsec:sdss_int_cor}.}
Then the observed MBH masses will be biased from the true ones by the systematic that is not well 
probed by the empirical estimator, i.e., 
\be
\ba
\log M_{\bullet,\rm obs}&=\log \bh+(a-\bar a)+(b-\bar b) \log L_\nu+s\\
&=\log \bh+b_{\bh} + \beta_{\rm L} \log (L_\nu/\overline{L}_\nu)+s,
\label{eq:mass_bias}
\ea
\ee
where $b_{\bh}=a-\bar a + \beta_{\rm L}\log \overline{L}_\nu$ describes the possible systematic error in the MBH mass estimator 
with a dependence on the QSO luminosity showing by the parameter $\beta_{\rm L}$ (similar to the parameter $\beta$ 
in \citet{Shen13} and \citet{Shen12}), and $\overline{L}_\nu$ is a characteristic luminosity. 

To cover such complexities, we assume that the PDF of the observational determined MBH masses ($M_{\bullet,\rm obs}$) around the true MBH mass ($M_\bullet$) is described by a Gaussian function
\be
\ba
&P(M_{\bullet, \rm obs}|\bh)=
\frac{1}{\sqrt{2\pi}\sigma_{M_\bullet}}\\
&\times\exp\left\{-\frac{[\log (\bh/M_{\bullet,\rm
obs})+b_{\bh}+\beta_{\rm L}\log (L_{\nu}/\overline{L}_{\nu})]^2}{2\sigma_{M_\bullet}^2}\right\}.
\label{eq:pselect}
\ea
\ee
Here we set $\log (\overline{ L}_\nu/$erg s$^{-1})=44.67$, which is the QSO luminosity when $\bh=10^8\msun$, $\log \lambda=-0.5$ and $\epsilon=0.1$.

We also assume that the selection function $\Omega(L_\opt,z)$ is simply given by
\be
\Omega(L_\nu,z)=\left\{\begin{array}{cc} \Omega_0, & L_\nu\ge L_{\rm lim}(z),\\
0, & L_\nu<L_{\rm lim}(z), \end{array}\right.
\label{eq:sel}
\ee
for a flux limit QSO survey.
Here $\Omega_0$ is the fraction of sky coverage of the survey, and $L_{\rm lim}(z)$ is the luminosity limit determined by the flux limit $i_{\rm lim}$ of the survey. 
For QSOs in SDSS Data Release 7, the effective sky coverage is about $6248$\,deg$^{2}$ \citep{Shen11}
and thus $\Omega_0 \simeq 0.15$ if the luminosity of a QSO at wavelength $2500$\AA\ is larger than the luminosity limit  $L_{\rm lim}(z)$. 
The fraction of QSOs with mass measurements in these spectroscopically targeted samples is close to unity, thus it can be safely ignored \citep[see discussions in][]{Schulze15}.
The $L_{\rm lim}$ at redshift $z$ for a given flux limit $i_{\rm lim}$ of the SDSS survey
can be found in \citet{Richards06}.

\subsection{The distribution functions of the intrinsic QSO properties}
\label{subsec:distribution}

\begin{figure}
\center
\includegraphics[scale=0.6]{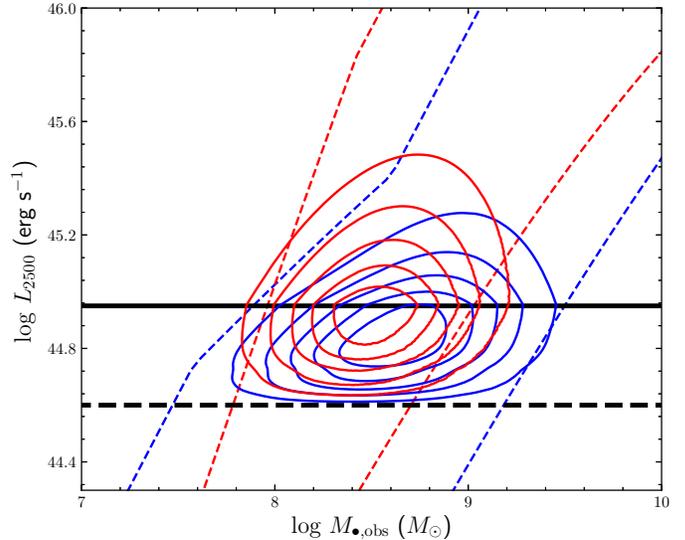}

\caption{
Distributions of mock QSOs at redshift $0.5<z<0.7$ on the plane of the optical luminosity versus the observational estimated MBH mass.
Color contours show the results obtained from a model by assuming an intrinsic relation $\epsilon \propto \bh^{\alpha_\epsilon}$ with $\alpha_\epsilon=-1$ (red contours) and $\alpha_\epsilon=1$ (blue contours), respectively. The contour levels from inside to outside represent those with relative number density per pixel values of {
	$5/6, 2/3, 1/2, 1/3,$ and  $1/6$ of peak value}, respectively.
The red and blue dash lines show the corresponding optical luminosity when $\lambda=1$ (left) or $\lambda=0.01$ (right)
{
	(For these lines the horizontal axis is $\log \bh$ ($\msun$))}.
 For details of the models that are used to obtain the mock samples, see Section~\ref{sec:mock_test}.  The horizontal black solid (or dash) line indicates the luminosity limit of $\log (L_{2500}/$erg s$^{-1})=44.9$ (or $44.6$) for QSOs at $z=0.7$ (or $0.5$), which are estimated by adopting the magnitude limit of the SDSS QSOs surveys, i.e., $i$ band limit of $19.1$\,mag.
}
\label{fig:ill_mobs_lopt}
\end{figure}

We assume that the intrinsic PDF of $\bh$ and $\lambda$ for QSOs within a sufficiently small redshift bin ($z$ to $z+dz$) can be approximated as
\be
\Phi(\bh,\lambda,z) = \Psi(\bh, \lambda) \Phi_z(z)
\label{eq:phibhGamma}
\ee
where $\Phi_z(z)$ represents the redshift evolution of the normalization and $\Psi_(M_\bullet,\lambda)$ represents the joint distribution of MBH mass and Eddington ratio.
For simplicity, we assume
\be
\Phi_z(z)=10^{\gamma_z(z-z_0)},
\label{eq:phiz}
\ee
where $\gamma_z$ and $z_0$ are two free parameters.  Below we fix $z_0=0.6$ simply because it is close to the middle of the redshift range that is considered in this paper.

We further set
\be
\Psi(\bh, \lambda)= \Phi_{M}(\bh)\Phi_\lambda(\lambda|\bh),
\ee
where $\Phi_M(\bh)$ is described by a Schechter-like function
\footnote{We have also try an alternative double power-law model:
$$
\Phi_{M}(\bh)=\frac{\Psi_\star}{\log_{10} e}\left( \frac{\bh}{M_\star}\right )^{\alpha_M+1} \left[1+\left(\frac{\bh}{M_\star}\right)^2\right]^{\frac{ \beta_M-\alpha_M}{2}},
%
%\label{eq:phimdp}
%
$$
However, our following calculations show that there is no clear difference between the results obtained by adopting these two different distribution forms. In the rest of the paper we discuss
only the results from the Schechter-like function.
}
\be
\Phi_{M}(\bh)=\frac{\Psi_\star}{\log_{10} e}\left( \frac{\bh}{M_\star}\right )^{\alpha_M+1} \exp\left( -\left[\frac{\bh}{M_\star}\right]^{\beta_M}\right),
\label{eq:phim}
\ee
where $\Psi_\star$, $M_\star$, $\alpha_M$, and $\beta_M$ are all free parameters.

The conditional Eddington ratio distribution for QSOs with the same $\bh$ is assumed to be described by
\be
\ba
\Phi_\lambda(\lambda|\bh)&=\frac{1}{\log_{10} e} \left[ \frac{\lambda}{\lambda_*(\bh)} \right]^{\alpha_{\lambda}+1} \exp\left\{-\left[\frac{\lambda}{\lambda_*(\bh)}\right]\right\}, \\
\label{eq:philambda}
\ea
\ee
%{\color{magenta}
	or alternative a Gaussian function
\be
\ba
\Phi_{\lambda}(\lambda|\bh)&=\frac{1}{\log_{10} e\sqrt{2\pi} \alpha_\lambda}
\exp \left[-\frac{(\log \lambda-\log \lambda_*(\bh))^2}{2\alpha_\lambda^2}\right]
\label{eq:philambdadp_gauss}
\ea
\ee
%}
%
where
\be
\log \lambda_*(\bh)=\log \lambda_{0}+k_\lambda(\log \bh-8),
\ee

$\lambda_0$, $k_\lambda$, and $\alpha_\lambda$ are all free parameters.

We assume that the intrinsic $\epsilon-\bh$ relation is simply described by
\be
\epsilon =
\begin{cases}
c_\epsilon M_8^{\alpha_\epsilon}, & \quad \\ %\text{otherwise},
0.038, & \quad \text{if } c_\epsilon M_8^{\alpha_\epsilon} <0.038, \\
0.42,   & \quad \text{if } c_\epsilon M_8^{\alpha_\epsilon} > 0.42,
\end{cases}
\label{eq:epsilonM}
\ee
where $c_\epsilon$ and $\alpha_\epsilon$ are two free parameters, $M_8=\bh/10^8 \msun$. The MBH spin is in the range from $-1$ to $1$, and correspondingly $\epsilon$ is confined in the range from $0.038$ to $0.42$ in the standard thin disk accretion model.
In the model, the inclination angle $i$ for each QSO is fixed at $\arccos 0.8$.

As seen from the above settings, the maximum likelihood model contains in total of $13$ free parameters, i.e., $(c_\epsilon, \alpha_\epsilon, \Psi_\star, M_\star, \alpha_M, \beta_M, \alpha_\lambda, \lambda_{0}, k_\lambda, \sigma_{M_{\bullet}}, b_{M_{\bullet}}, \beta_{\rm L},\gamma_z)$. The normalization of the likelihood function, i.e., Equation~\eqref{eq:maxim_likelihood}, depends on the assumed range of $\bh$ and $\lambda$. We assume that $0.01<\lambda<1$ and $6.5<\log (\bh/M_\odot)<10.5$, which cover almost all of the SDSS QSOs at $z<0.9$ (see also Fig.~\ref{fig:mobs_lopt}).

\section{Mock sample test of the method}
\label{sec:mock_test}

In this section, we generate mock QSO samples by setting different sets of the model parameters and illustrate how the observed PDFs are affected by the assumed
$\epsilon-\bh$ relation. We further use the maximum likelihood method described in Section~\ref{sec:maxlkhood} to check how well the $\epsilon-\bh$ relation can be reconstructed from the mock observations, or how it is biased due to a number of 
complexities (e.g., the host galaxy contamination, or scatterings in the inclinations).
The details are described in the following sections.

\subsection{Mock QSOs}
\label{subsec:mock_sample}

\begin{figure}
\center
\includegraphics[scale=0.7]{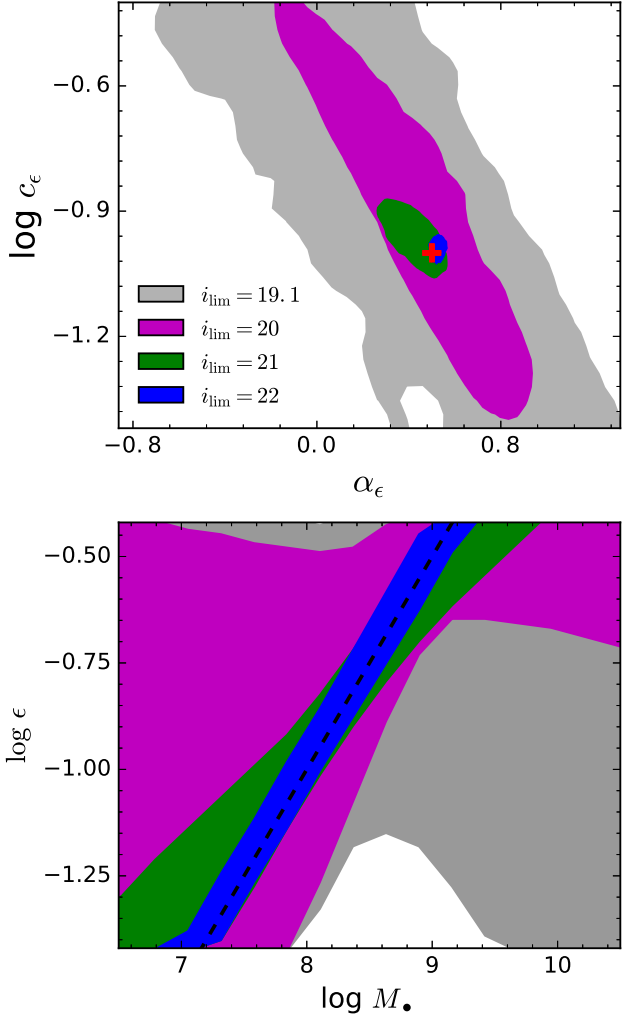}

\caption{
Constraints on the $\epsilon-\bh$ relation obtained for a group of mock samples with different magnitude limits ($i_{\rm lim}$) and intrinsic $\epsilon-\bh$ relations by using the maximum likelihood method. The results are obtained by allowing $11$ of those $13$ model parameters (as described in the last paragraph of Sec.~\ref{subsec:distribution}) to vary freely, but the rest two parameters, i.e., $b_{M_{\bullet}}$ and $\beta_{\rm L}$, fixing to zero.
The magnitude limits are adopted as $i_{\rm lim}=19.1$, $20$, $21$, and $22$, respectively, with the first one is the same as that for the SDSS samples at redshift $z<2.9$.
Top panel: color contours show the model results of the $2\sigma$ confidence regions of the marginalized likelihood function for $\alpha_\epsilon$ and $\log$ $c_\epsilon$. The red cross marks the input values of $\alpha_\epsilon=0.5$ and $\log$ $c_\epsilon=-1$). Bottom panel: the two dimensional $2\sigma$ 
confidence contour of the intrinsic $\epsilon-\bh$ relation. The black dash line marks the input $\epsilon-\bh$ relation.
}
\label{fig:mcmc_con}
\end{figure}

Here we introduce the initial conditions and detailed implementations for the mock sample generation. If not otherwise stated, these conditions and implementations are also used for Section~\ref{subsec:flux_limit}, \ref{subsec:scatter_cosi_epsilon}, \ref{subsec:host_galaxy_con}. We assume $\log \Psi_\star =-5.0$, $\log M_\star= 7.5$, $\alpha_M= -1.3$, $\beta_M =0.5$, $\alpha_\lambda=-1.5$, $\log \lambda_{0}=-1$, $k_\lambda=0.2$, and $\gamma_z=0$, such that they are close to the fitting results from \citet{Schulze15} for SDSS QSOs at $z=0.6$. Then the joint PDF $\Phi(\bh, \lambda, z)$ can be obtained from Equations~(\ref{eq:phiz})-(\ref{eq:philambda}). Given the above initial conditions, the total number of the intrinsic samples $N_{\rm int}$ can be obtained by
\be
N_{\rm int}=\iiint \Phi(\bh, \lambda,z) \frac{dV}{dz} dzd\bh d\lambda.
\ee
We generate a number of $N_{\rm int}$ mock QSOs by the Monte-Calro method. For simplicity, we also assume $f_{\rm col}=1.7$ and 
$\cos i =0.8$, if not otherwise stated. The radiative efficiency $\epsilon$ of each mock QSO is given by Equation~\eqref{eq:epsilonM}, by assuming $c_\epsilon = 0.1$ and $\alpha_\epsilon$ adopts a value in between $-1$ and $1$. 
Then the optical band luminosity $L_\nu$ at $2500$\AA~ of each mock QSO is estimated according to Equation~\eqref{eq:loptad}. 

We find that the constraint on the intrinsic $\epsilon-\bh$ relation weakly depend on the sample size or correspondingly 
$\Omega_0$ for a given magnitude limit. Therefore, we simply assume that $\Omega_0=1$.
We assume that a QSO at redshift $z$ is ``observable'' if its $L_{\nu}\ge L_{\rm lim}(z)$, 
and then the ``observed'' mass of this mock QSO is set according to Equation~\eqref{eq:pselect} 
by assuming $\sigma_{\bh}=0.3$, $\beta_{\rm L}=0$ and $b_{\bh}=0$. Finally, we obtain a sample of $N$ mock QSOs with ``observational'' properties of $M_{\bullet,{\rm obs},i}$, $L_{\nu,i}$, and $z_i$, $i=1,\cdots, N$. According to such a mock sample, we can then use the maximum likelihood in Equation~\eqref{eq:maxim_likelihood} to recover the input parameters, and check how well can they be reproduced.

Figure~\ref{fig:ill_mobs_lopt} shows the distributions of mock QSOs on the $L_{\nu}-M_{\bullet,\rm obs}$ plane resulting from the same $\Phi(\bh,\lambda, z)$ but two  different $\epsilon-\bh$ relation, one with $\alpha_\epsilon =1$ (blue contours) and the other with  $\alpha_\epsilon =-1$ (red contours). These two distributions are quite different, suggesting that the apparent $L_\nu-\bh$ joint distribution depend not only on $\Phi(\bh,\lambda,z)$ but also on the intrinsic $\epsilon-\bh$ correlation, if any. Due to such differences, the intrinsic $\epsilon-\bh$ relation can be extracted from the 2D distribution of QSOs on the luminosity versus observed MBH mass plane.

\subsection{Effects of the flux limit of the QSO survey}
\label{subsec:flux_limit}

\begin{figure}
\center
\includegraphics[scale=0.7]{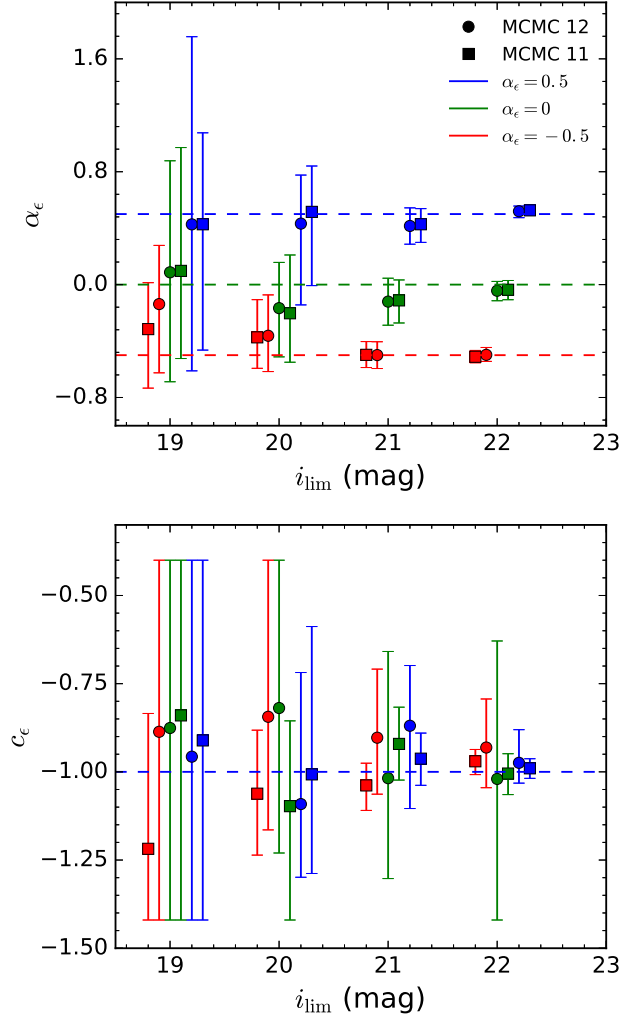}

\caption{
Constraints on $\alpha_\epsilon$ (top panel) and $c_\epsilon$ (bottom panel) by using the maximum likelihood method to fit a group of mock QSO samples with different assumed magnitude limits $i_{\rm lim}$. The settings of the simulations are similar to those in Fig.~\ref{fig:mcmc_con}. The blue, green, and red symbols (and their associated $2\sigma$ errorbars) in the top (bottom) panel represent the constraint on $\alpha_\epsilon$ ($c_\epsilon$) obtained from the MCMC calculations for the cases with different survey magnitude limit ($i_{\rm lim}=19.1$, $20$, $21$, and $22$\,mag, respectively) and different input $(\alpha_\epsilon,c_\epsilon)= (0.5,0.1)$, $(0,0.1)$, and $(-0.5,0.1)$, respectively. Note that for clarity, the positions of each symbol slightly offset in $x$-axis direction from each given $i$ magnitude limit. Filled circles and solid squares represent the constraints obtained from the MCMC calculations by utilizing the $11$-parameter and $12$-parameter models, respectively. This figure clearly shows that the $\epsilon-\bh$ relation can be well reconstructed if the magnitude limit of a survey can reach $i_{\rm lim} \gtrsim 20-21$.
}
\label{fig:mcmc_eff}
\end{figure}

\begin{figure*}
\includegraphics[scale=0.7]{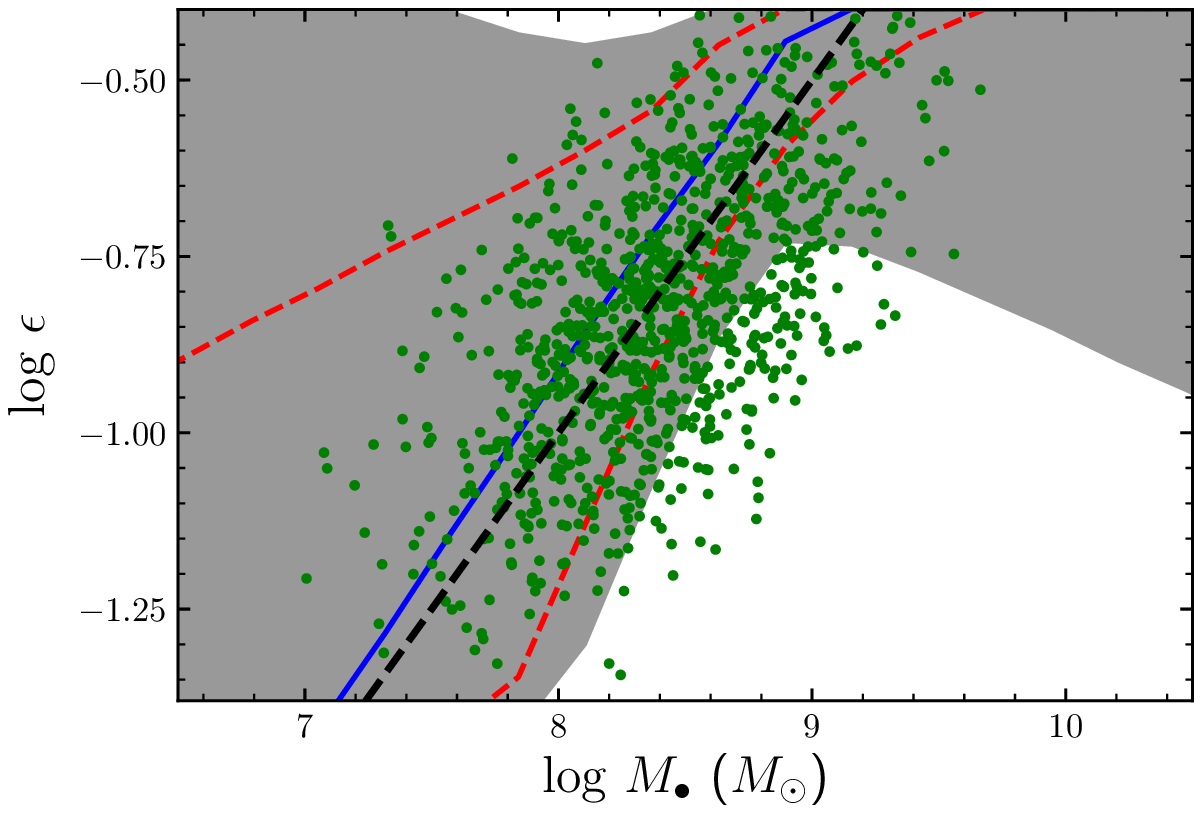}
\includegraphics[scale=0.7]{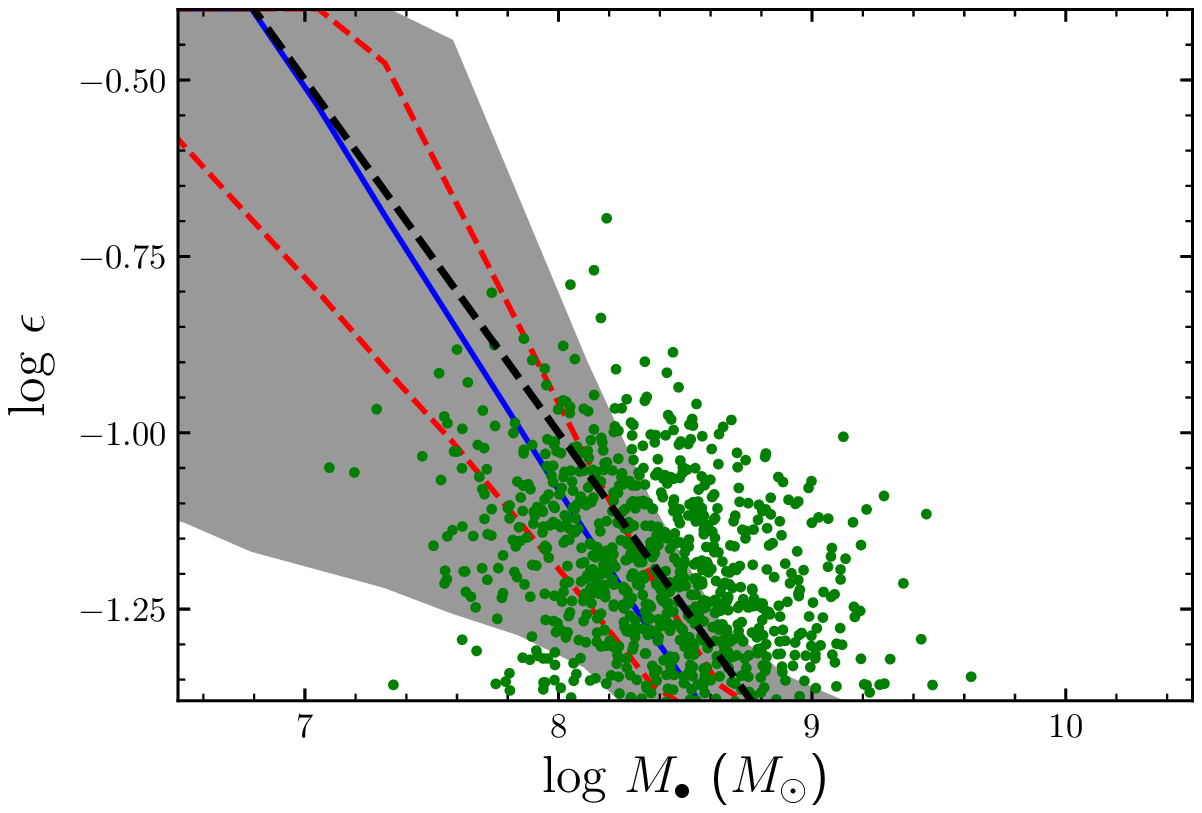}
\caption{
The two dimensional confidence level constrained for the intrinsic $\epsilon-\bh$ relation, when scattering of $\cos i$ and Eq.~\eqref{eq:epsilonM} in the mock samples are considered. The green dots show the input value of $\epsilon$ of $1000$ observed samples (partially selected, for clarity). The regions enclosed by the red dash lines show the $1\sigma$ confidence level of the constraints, and the grey shadows show the $2\sigma$ confidence level. In each panel, the blue solid line shows the best-fit, which corresponds to the median value of the marginalized likelihood function, and the black dash line marks the input 
$\epsilon-\bh$ relation.
}
\label{fig:int_eff}
\end{figure*}

The distribution of QSOs from a survey on the $L_\nu-M_{\bullet, \rm obs}$ plane only shows those QSOs above the magnitude/flux limit of the survey (see Figure~\ref{fig:ill_mobs_lopt}). Therefore, the constraints on the $\epsilon-M_{\bullet}$ relation obtained from such a distribution may strongly depend on the flux/magnitude limit of the survey. 

To investigate how the constraints on the model parameters depend on the flux limit, we perform a number of Monte Carlo Markov Chain (MCMC) calculations. We consider three cases, i.e., $\alpha_\epsilon$ is assumed to be $-0.5$, $0$, and $0.5$, respectively. For each of these cases, we first generate mock QSO samples at the redshift bin $0.5 <z <0.7$\footnote{It can be similarly done for any other redshift bins. }
and the $i$-band magnitude limit is set to be $i_{\rm lim} = 19.1$\,mag (similar to the SDSS QSO survey), $20$, $21$, or $22$\,mag.
Then we use the MCMC method to recover the input values of those model parameters by adopting the likelihood function given by Equation~\eqref{eq:maxim_likelihood}. The model considered here contains $11$ parameters (with $b_{M_\bullet} =0$ and $\beta_{\rm L}=0$), or a model with $12$ parameters by only fixing $\beta_{\rm L}=0$, so that we can see the effects of constant 
mass bias $b_{M_\bullet}$ by comparing these two models. 

To show more clearly the constraints of the intrinsic $\epsilon-\bh$ relation, we obtain the two dimensional (2D) confidence level for the reconstructed $\epsilon-\bh$ relation according to the accepted values of $\alpha_\epsilon$ and $c_\epsilon$ in each chain of the MCMC fittings. The results for the case $\alpha=0.5$ are shown in Figure~\ref{fig:mcmc_con}. 
We can see that the input $\epsilon-M_{\bullet}$ relation can be well recovered when the flux limit is sufficiently low, i.e., $i_{\rm lim} \ga 20-21$ mag. Note that the best-fit values of $\alpha_\epsilon$ and $c_\epsilon$ slightly deviate from the input ones when $i_{\rm lim}=22$, which is simply due to the random noises introduced by the Monte Carlo method in generating the mock QSO samples.

When $i_{\rm lim}\la 19$\,mag, there is a strong degeneracy between $\alpha_\epsilon$ and $c_\epsilon$ (top panel of Fig.~\ref{fig:mcmc_con}) and the $\epsilon-\bh$ relation cannot be well constrained (bottom panel of Fig.~\ref{fig:mcmc_con}). Such a degeneracy is expected as low mass QSOs can not be observed, and the observed sample covers only a small range of $\bh$. 
If these faint QSOs can be included in the sample, as illustrated in Figure~\ref{fig:mcmc_con} when $i_{\rm lim} \ga 20-21$\,mag, the degeneracy will then be almost disappeared.

Figure~\ref{fig:mcmc_eff} shows the constraints on $\epsilon-M_\bullet$ relation for cases with different input values for 
$\alpha_\epsilon$ by using either the $11$ or $12$ parameter model.
We can see that, in most cases, the constraints on both $\alpha_\epsilon$ and $c_\epsilon$ are significantly improved 
if the magnitude limit of the QSO survey increases to $\ga 20-21$\,mag. Note here that the value of $c_\epsilon$ is always poorly constrained 
(see bottom panel of Fig.~\ref{fig:mcmc_eff}) if there is no intrinsic $\epsilon-\bh$ relation 
(i.e., $\alpha_\epsilon \sim 0$) and if the $b_{M_\bullet}$ is included in the MCMC fitting.
As $b_{M_\bullet}$ reflects the systematic errors in the mass estimates, $c_\epsilon$ is strongly degenerate 
with $b_{M_\bullet}$ when $\alpha_\epsilon \sim 0$. However, such a degeneracy can be breakup when 
$\alpha_\epsilon$ is significantly different from $0$.

We find that the constraints on the intrinsic $\epsilon-\bh$ relation depend also on the initial value of $\alpha_\epsilon$. If $\alpha_\epsilon <0$, the QSOs in the mock samples are statistically brighter than those if $\alpha_\epsilon >0$ (See Fig.~\ref{fig:ill_mobs_lopt}), thus the constraints are slightly stronger (See Figure~\ref{fig:mcmc_eff}).
For other parameters in the model, we also find that they can be reproduced with considerable accuracy. 

Similar constraints on the intrinsic $\epsilon-\bh$ relation can be also obtained for mock samples at other redshift bins if assuming that the intrinsic distribution function $\Phi(\bh,L_\nu,z)$ depends weakly on the redshift. According to the relation between the luminosity and the $i$-band magnitude limit given by~\citet{Richards06},  $i_{\rm lim}= 20 \sim21$\,mag at $z\sim0.6$ corresponds to $L_\nu > 10^{44.4}-10^{44}{\rm erg\,s^{-1}}$ and this luminosity range corresponds to $i_{\rm lim} \sim 19-20$ mag at $z=0.4$. Therefore, it is possible to get a reasonably constraints by using the SDSS QSO sample at redshift $z\la 0.4$ as its magnitude limit is $19.1$\,mag. Similarly, if a QSO survey can have the magnitude limit of $i_{\rm lim}=22$\,mag, then it would provide strong constraints on the $\epsilon-\bh$ relation at redshift $z\la 1.5$.

\subsection{Scattering of $\cos i$ and $\epsilon$}
\label{subsec:scatter_cosi_epsilon}

\begin{figure}
\center
\includegraphics[scale=0.85]{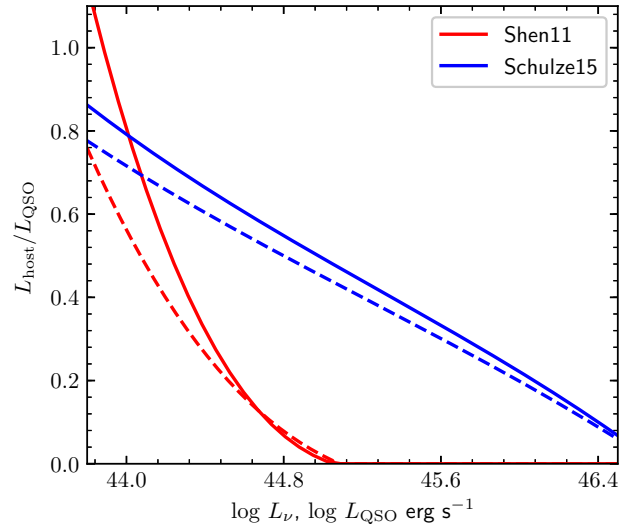}
\caption{
Host galaxy to QSO luminosity ratio, illustrating the importance of host galaxy contamination. Blue and red lines indicate the ratio given by \citet{Shen11} and \citet{Schulze15}, respectively. The solid and dash lines represent this ratio as functions of the total luminosity $L_\nu$ and the QSO luminosity $L_{\rm QSO}$, respectively.
}
\label{fig:host}
\end{figure}

For the mock simulations in the previous section, we have set $\cos i=0.8$ and assume that the $\epsilon-\bh$ relation is explicitly given by Equation~\eqref{eq:epsilonM}. In reality, however, it is likely that $\cos i$ is randomly distributed over a range of values, and the $\epsilon-\bh$ relation may have some scatters.  To check whether the maximum likelihood method can still work well under these circumstances, we generate a group of mock samples with settings similar to those in Section~\ref{subsec:mock_sample} except $\cos i$ uniformly distributed between $(0.6, 1.0)$ and a Gaussian scatter of $0.1$\,dex on the $\epsilon$ given by Equation~\eqref{eq:epsilonM}. We generate mock QSO samples at $0.3<z<0.5$ and set $i_{\rm lim}=19.1$\,mag to mimic the case of SDSS DR7 QSOs. We then use the MCMC fitting technique to obtain constraints on the $\epsilon-\bh$ relation.

Figure~\ref{fig:int_eff} shows the results of the constraints on the intrinsic relation.  The green dots show the exemplified value of $\epsilon$ for the `observed' sample. The blue solid line represents the best-fit value and the region enclosed by the gray shadows shows the $2-\sigma$ constraints. It can be seen that the $\epsilon-\bh$ relation can still be well reproduced, even if there are some scatters in the radiative efficiencies. As demonstrated in Section~\ref{subsec:flux_limit}, the constraints on the intrinsic correlations is stronger for the samples with $\alpha_\epsilon<0$ than those with $\alpha_\epsilon >0$.

If the scatter is large, then Equation~\eqref{eq:epsilonM} is not so meaningful and the reconstructed value of $\alpha_\epsilon$ will then likely be zero, which may only reflect an effective intrinsic relation. We find that the other model parameters, defining the form of $\Phi(\bh,\lambda,z)$, can also be well reconstructed within $3\sigma$ confidence level.
Therefore, we conclude that the maximum likelihood method introduced in this paper can work well even if there is a large scatter in the $\epsilon-\bh$ relation.  Note here that there might be a scatter for the intrinsic $\epsilon-\bh$ relation, if any, which is not considered above. In principle, one may add more model parameters to include such a possiblity for the $\epsilon-\bh$ relation when considering the likelihood function. However, the calculations will be more complicated and computationally heavy\footnote{The scatterings of $\epsilon$ can be described by a probability function $\Phi_\epsilon(\epsilon|\bh)$ similar to that shown in Equations~\eqref{eq:philambda} and \eqref{eq:philambdadp_gauss}, then the mass-luminosity joint distribution is given by 
$$
\ba
\Phi(\bh, L_{\nu})&=\frac{\partial}{\partial L_\nu}\int^{L_\nu}\Phi(\bh, L_{\nu}')dL_\nu'\\
&=\frac{\partial}{\partial L_\nu}
\left[~~~\iint\limits_{L_\nu(\lambda, \epsilon)<L_{\nu}} \Phi(\bh, \lambda)\Phi_\epsilon(\epsilon|\bh)d\lambda d\epsilon\right],
\ea
\label{eq:cumulative_lnu}
$$
where $L_{\nu}(\lambda, \epsilon)$ is given by Equation~\eqref{eq:loptad}.
%Note that Equation~\eqref{eq:phiomobs} should also now use $\Phi(\bh, L_{\nu})$ in the above equation.
Then the likelihood function of Equation~\eqref{eq:maxim_likelihood} can be accordingly modified (by using the integrals of the above equation) and calculated. }. In this paper, we do not intend to introduce such model parameters.

\subsection{Effect of host galaxy contamination}
\label{subsec:host_galaxy_con}

The contamination from host galaxies makes some of those QSOs with intrinsical luminosities below the magnitude limit become observable, and thus leads to a modification of the distribution of those QSOs on the observed $\bh-L_\nu$ plane.
Therefore, the recovered intrinsic $\epsilon-\bh$ relation may be biased if the host galaxy contamination is not well modeled for the QSO sample(s).

Here we use MCMC simulations to investigate how the constraints of the intrinsic correlations are effected by the uncorrected
host galaxy contamination.
For the demonstration purpose, here we consider two different models for host contamination. The first one is from \citet{Shen11}, in which the ratio of host to QSO luminosity at a given QSO luminosity is given by\footnote{Note that this correction is for the $5100$\AA\ luminosity. However, we adopt it here only to illustrate the importance of the host galaxy contamination. The correction in \citet{Shen11} is a function of the total luminosity (see Eq.~\eqref{eq:shen11_bw}), which is not initially known for the mock QSOs, thus, we need to convert it as a function of intrinsic QSO luminosity by refitting to the original equation.}
\be
\frac{L_{\rm host}}{L_{\rm QSO}}=0.5621-0.8971x+0.4104x^2-0.05620x^3,
\label{eq:shen11_fw}
\ee
where $x=\log (L_{\rm QSO}/{\rm erg~s}^{-1})-44$, and $x+44<45.03$.
Another one is derived from~\citet{Schulze15}(See their Fig.~2):
\be
\frac{L_{\rm host}}{L_{\rm QSO}}=0.7158-0.2941x+0.03812x^2-0.01021x^3.
\label{eq:schulze15_fw}
\ee
and
%$\log (L_{\rm QSO}/{\rm erg~s}^{-1})<46.67$.
$x+44<46.67$. The upper limits of $x$ for the above two equation are obtained by setting the host to QSO luminosity ratio to zero. For QSOs with luminosities larger than the upper limit, the host contamination is negligible. Figure~\ref{fig:host} shows the ratio of host to QSO luminosity given by these two models.

We generate a group of samples similar to those in Section~\ref{subsec:mock_sample}
but at lower redshift $0.3<z<0.5$ (such that the host galaxy contamination becomes
more important). Initially we set $\alpha_\epsilon=-0.5$ (or $\alpha_\epsilon=0.5$).
We then add the luminosity of galaxies into the samples according to Equation~\eqref{eq:shen11_fw} or \eqref{eq:schulze15_fw}. 
For each group of samples we use MCMC simulations (of which the host galaxy contamination is not included), 
to investigate whether the $\epsilon-\bh$ correlations can be correctly recovered.

The results are as follows. we found that if the host galaxy contamination is similar to \citet{Shen11}, the recovered 
$\alpha_\epsilon$  will be biased to higher values. 
For example, the recovered $\alpha_\epsilon$ is $\sim0.2$ (or $0.7$) if the intrinsic $\alpha_\epsilon=-0.5$  (or $0.5$). 
However, if the contamination is similar to that given by \citet{Schulze15}, the effect of host galaxy contamination is not important. Both the intrinsic $\epsilon-\bh$ relation and other model parameters can be well recovered.

These results suggest that whether the intrinsic $\epsilon-\bh$ relation can be well recovered or not depends on
the exact form of the host galaxy contamination. For example, the host galaxy contamination from 
\citet{Shen11} is more severe than those in \citet{Schulze15} for faint QSOs (See in Figure~\ref{fig:host}).
The $\bh-L_{\nu}$ distribution of the observed samples are more significantly modified, resulting in a strong bias 
of the intrinsic correlations. The above results suggests that, for a comprehensive study, it is important to 
investigate the effects of host galaxy contamination in the recovery of $\epsilon-\bh$ correlations, especially
for low redshift QSO samples.

\section{Application to the SDSS DR7 QSO samples}
\label{sec:applying_to_sdss}

\begin{table*}
\begin{center}
\caption{
	Different models and its settings considered in this paper.
}
\begin{tabular}{ccccccccccccc} \hline \hline
Name & fixed parameters & Host Galaxy correction &
$f_{\rm col}$ & BHMF & ERDF \\ \hline
M11N2S & ${b_\bh=0}, \beta_{\rm L}=0$ &  None & $1.7$ & Schechter & Schechter\\
M12N2S & $\beta_{\rm L}=0$ &  None & $1.7$ & Schechter & Schechter \\
M12N1S & $\beta_{\rm L}=0$ &  None & $1$ & Schechter & Schechter\\
M12S2S & $\beta_{\rm L}=0$ &  Shen11 & $1.7$ & Schechter & Schechter\\
M12U2S & $\beta_{\rm L}=0$ &  Schulze15 & $1.7$ & Schechter & Schechter\\
%
%M13N2D & None &  None  & $1.7$ & Double power & Schechter \\
%
M13N2S & None &  None & $1.7$ & Schechter & Schechter\\
%
%M13S2S & None &  Shen11 & $1.7$ & Schechter & Schechter\\
%
M11N2SG & ${b_\bh=0},\beta_{\rm L}=0$ &  None & $1.7$ & Schechter & {Gaussian}\\
M13N2SG & None &  None & $1.7$ & Schechter & {Gaussian}\\
 \hline \hline
\end{tabular}
\label{tab:model}
\end{center}
\tablecomments{Here {$b_\bh$} and $\beta_{\rm L}$ are two parameters describing the luminosity-dependent bias of the QSO mass estimator (defined in Eq.~\ref{eq:mass_bias}). Host galaxy contamination correction is given by either Equation~\eqref{eq:shen11_bw} (``Shen11'') or Equation~\eqref{eq:schulze15_bw} (``Schulze15''), respectively.
}
\end{table*}

\begin{table*}
	\begin{center}
\caption{
	Constraints on $\alpha_\epsilon$ and $c_\epsilon$ obtained from the fitting of SDSS QSOs in different redshift bins to those models listed in Table~\ref{tab:model}.
}

\begin{tabular}{ccccccccccccccc} \hline
\multirow{2}{*}{Model} &
\multicolumn{2}{c}{z=0.34-0.5(H$\beta$)} & & \multicolumn{2}{c}{z=0.5-0.7(H$\beta$)}
&& \multicolumn{2}{c}{z=0.5-0.7(MgII)} & & \multicolumn{2}{c}{z=0.7-0.9(MgII)}
\\
   \cline{2-3} \cline{5-6} \cline{8-9} \cline{11-12}
   &
  $\alpha_\epsilon$       & $c_\epsilon$
& & $\alpha_\epsilon$       & $c_\epsilon$
&  & $\alpha_\epsilon$       & $c_\epsilon$          \\
  \hline
M11N2S  & $ 0.75^{+0.32}_{-0.56} $  & $-1.13^{+0.45}_{-0.29} $& &
$ 0.54^{+0.26}_{-0.27} $  & $-1.11^{+0.25}_{-0.30} $ & &
$ -0.50^{+0.41}_{-0.50} $ & $-1.01^{+0.39}_{-0.41} $ & &
$ 0.63^{+1.08}_{-1.09} $ & $-0.89^{+0.49}_{-0.53} $     \\
M12N2S   & $ 0.9^{+0.13}_{-0.25} $  & $-1.38^{+0.10}_{-0.04} $& &
$ 0.65^{+0.15}_{-0.30} $  & $-1.32^{+0.18}_{-0.09}$ & &
$ -0.74^{+0.54}_{-0.40} $  & $-0.66^{+0.26}_{-0.36}$ & &
$ 0.02^{+0.68}_{-0.61} $  & $-0.95^{+0.55}_{-0.47}$
\\
M12N1S   & $ 1.13_{-0.27}^{+0.15} $  & $-1.35_{-0.07}^{+0.11}$& &
$ 0.55_{-1.00}^{+0.30} $  & $-1.31^{+0.11}_{-0.17} $ & &
$ -0.36^{+0.51}_{-0.45} $  & $-0.65^{+0.25}_{-0.40} $ & &
$ 0.13_{-0.52}^{+0.47} $  & $-0.94^{+0.54}_{-0.48} $
\\
M12S2S   &  $ 1.06^{+0.13}_{-0.22} $  & $-1.39^{+0.09}_{-0.03} $& &
$0.67^{+0.2}_{-1.00} $  & $-1.38^{+0.07}_{-0.04} $ & &
$ -1.03^{+0.67}_{-0.49} $ & $-0.67^{+0.27}_{-0.34}$ & &
$ 0.09^{+0.72}_{-0.71} $ & $-0.93^{+0.53}_{-0.49} $
\\
M12U2S   & $ 0.87^{+0.42}_{-1.47} $  & $-1.27^{+0.18}_{-0.15} $& &
$ 0.77^{+0.15}_{-0.20} $  & $-1.33^{+0.14}_{-0.09} $ & &
$ -0.68^{+0.88}_{-0.60} $ & $-0.70^{+0.30}_{-0.51} $ & &
$ -0.08^{+0.67}_{-0.68} $ & $-0.91^{+0.51}_{-0.51} $
\\
%M13N2D    & $ 0.57^{+0.12}_{-0.15} $  & $-1.04^{+0.13}_{-0.10} $& &
%$ 0.32^{+0.38}_{-0.89} $  & $-1.23^{+0.35}_{-0.19} $ & &
%$ -0.75^{+1.00}_{-0.56} $ & $-0.66^{+0.26}_{-0.35} $ & &
%$ -0.77^{+0.98}_{-0.73} $ & $-0.73^{+0.33}_{-0.49} $ \\
M13N2S   & $ 0.55^{+0.36}_{-0.22} $  & $-1.03^{+0.21}_{-0.39} $& &
$ 0.52^{+0.25}_{-0.43} $  & $-1.26^{+0.34}_{-0.16} $ & &
$ -0.62^{+0.88}_{-0.99} $ & $-0.73^{+0.33}_{-0.44} $ & &
$ -0.14^{+0.60}_{-0.42} $ & $-1.00^{+0.60}_{-0.42} $  \\
%
%M13S2S   & $ 0.99^{+0.13}_{-0.21} $  & $-1.38^{+0.09}_{-0.04} $& &
%$ 0.54^{+0.25}_{-0.64} $  & $-1.37^{+0.06}_{-0.05} $ & &
%$ -0.98^{+0.73}_{-0.91} $ & $-0.76^{+0.36}_{-0.38} $ & &
%$ -0.11^{+0.64}_{-0.74} $ & $-0.86^{+0.46}_{-0.56} $ \\
M11N2SG  & $ 0.27^{+0.14}_{-0.38}$ & $-0.72^{+0.11}_{-0.09} $& &
$ 0.02^{+0.26}_{-0.24} $  & $-0.62^{+0.15}_{-0.16} $ & &
$ -0.09^{+1.17}_{-1.05} $ & $-0.82^{+0.42}_{-0.60} $ & &
$ -0.54^{+1.38}_{-0.81} $ & $-0.91^{+0.51}_{-0.51} $  \\
M13N2SG   & $ 0.53^{+0.45}_{-0.26} $  & $-0.96^{+0.28}_{-0.46} $& &
$ 0.32^{+0.37}_{-0.32} $  & $-1.07^{+0.67}_{-0.35} $ & &
$ -0.87^{+1.36}_{-1.09} $ & $-0.79^{+0.39}_{-0.63} $ & &
$ -0.33^{+0.83}_{-1.08} $ & $-0.91^{+0.51}_{-0.51} $ \\
\hline
\end{tabular}
\label{tab:result}
\end{center}
\tablecomments{Here ``H$\beta$'' and ``MgII'' indicate those QSO samples in which the black hole masses are obtained by using the H$\beta$ and MgII mass estimators (``Schulze15''), respectively. The subscript and superscript associated with each of the best-fit values make the $2\sigma$ uncertainty range.
}
\end{table*}

In this section, we apply our maximum likelihood method to the SDSS QSOs samples to extract the underlying intrinsic correlations.  We apply our method to fit the QSOs in each redshift bins by varies different assumptions on the model (See Table~\ref{tab:model}). The details are in the following Sections.

\subsection{SDSS QSOs samples}

In this section, we briefly describe the SDSS QSO samples \citep{Shen11}. In the catalogue of SDSS QSOs [Data Release 7 (DR7)], $104746$ QSOs have $i$-band magnitude $M_i<-22$, at least one broad emission line broader than $1000\kms$, and the estimations of central MBH masses. About half of these SDSS QSOs (total $57959$) QSO at $0.3\le z\le 5$ were selected out to form a homogeneous, statistical sample, which is primarily a flux limited sample with $i$-band magnitude $m_i\le 19.1$ at $z\le2.9$, $m_i\le20.2$ at $z>2.9$ and an additional bright limit of $m_i\ge15$. We select the QSOs with $0.34 \le z\le0.9$ 
 in this sample (total number $15356$) to form our SDSS QSO samples.
 Among them, those QSOs with mass estimation errors $>0.5$\,dex were removed from the QSO sample. The total number of these QSOs is only $\sim3\%$ of the whole sample, thus the removal of them only slightly affects the overall completeness.

We have shown in Section~\ref{subsec:flux_limit} that the SDSS-like QSO samples beyond $z\sim0.6-0.8$ are unlikely set useful constraints on the intrinsic correlations, thus we do not include the QSO samples with $z>0.9$. Also, we notice that the completeness of the QSO samples at $z\le0.9$ are almost unity \citep{Richards06}, which greatly simplify our maximum likelihood method.
The $L_{2500}$ luminosity at $2500$\AA\ is converted from $M_i(z=2)$, the K-corrected $i$-band absolute magnitude \citep{Richards06}.

We divide those QSOs with $z<0.9$ into three redshift bins, i.e., $0.34<z<0.5$, $0.5<z<0.7$, and $0.7<z<0.9$. For QSOs with $0.34<z<0.7$, their masses can be obtained by utilizing the H$\beta$  estimator, while for those with $0.5<z<0.9$, their masses can be obtained by utilizing the MgII estimator. For the redshift bin $0.5<z<0.7$, QSO masses can be obtained by using either the H$\beta$ or MgII estimator. Therefore, a comparison between the results obtained by utilizing the H$\beta$ estimator and those by the MgII estimator in this bin may provide an estimation of the effects induced by possible systematical errors in different mass estimators. The total numbers of QSOs using the H$\beta$ estimator are $3420$ and $4025$ in the bins $0.34<z<0.5$ and $0.5<z<0.7$, respectively, while those using the MgII estimator are $4039$ and $3872$ in the bins $0.5<z<0.7$ and $0.7<z<0.9$, respectively. Figure~\ref{fig:mobs_lopt} shows the two dimensional number distribution of these QSOs in the $\log L_{2500}-\log M_{\bullet,\rm obs}$ plane.

\subsection{Models}

\begin{figure*}
\includegraphics[scale=0.7]{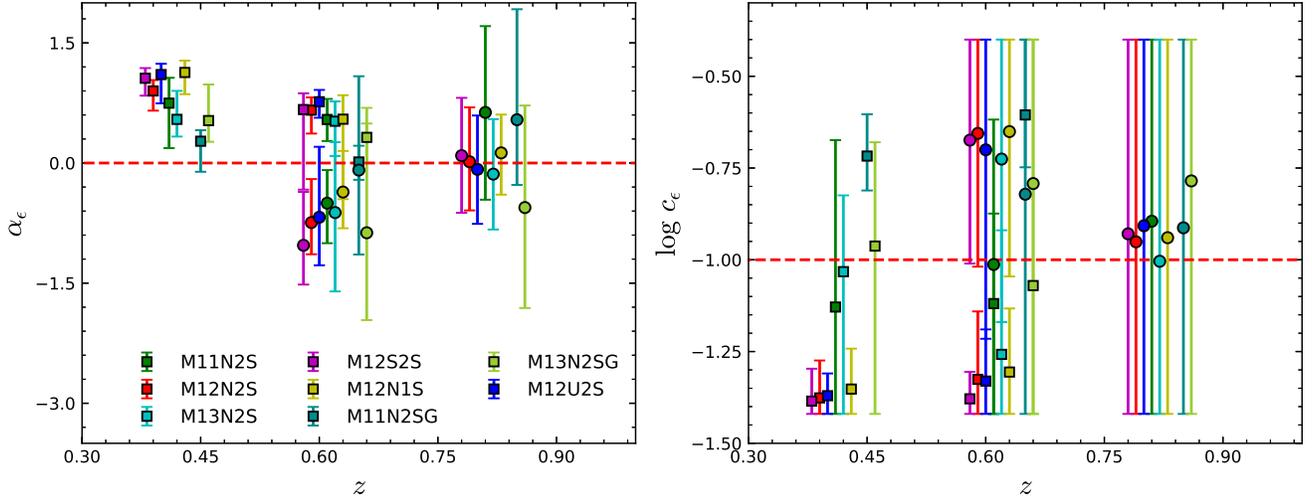}
\caption{
Constraints on $\alpha_\epsilon$ (left panel) and $c_\epsilon$ (right panel) obtained from the model fitting to the SDSS QSO samples at different redshift bins. Color symbols show the results obtained from those models listed in Table~\ref{tab:model} (as indicated by the text label in the figure) and the constrained $\alpha_\epsilon$ and $c_\epsilon$ values are listed in Table~\ref{tab:result}.
The error bars associated with each symbol indicate the $2\sigma$ confidence level. The red dash lines show the reference values of $\alpha_\epsilon=0$ and $c_\epsilon=-1.0$, respectively. Note that the masses of SDSS QSOs are obtained by adopting the H$\beta$ mass estimator in the redshift bin $0.34<z<0.7$ (filled squares) and the MgII mass estimator in redshift bing $0.5<z<1.5$ (filled circles). 
Note that the results obtained from different models at each redshift bin are slightly offset from each other in the horizontal direction, for the clarity of figure.
}
\label{fig:sdss_epsilon}
\end{figure*}

\begin{figure*}
\centering
\includegraphics[scale=0.6]{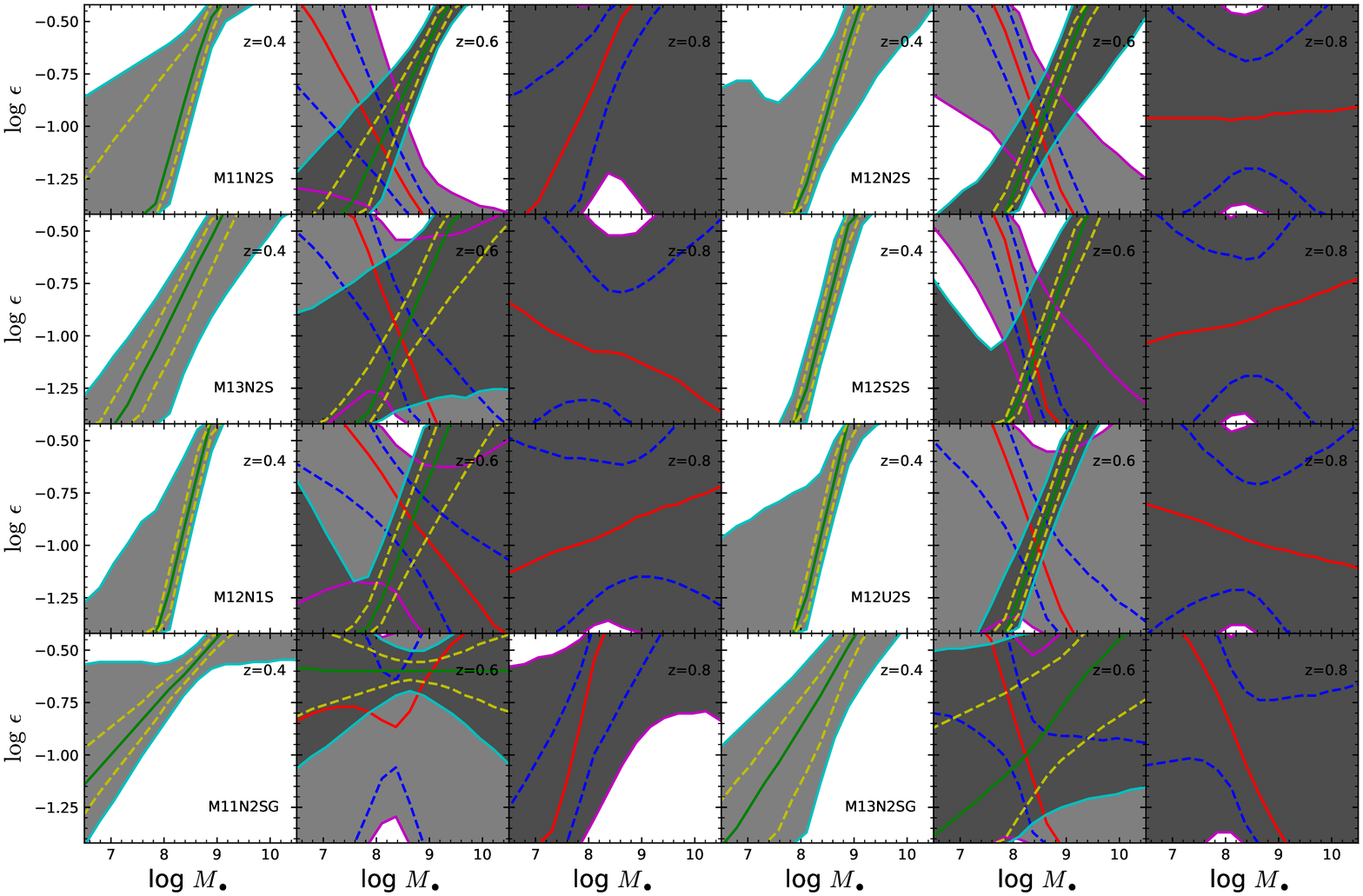}
\caption{
Two dimensional constraints on the intrinsic $\epsilon-\bh$ relation obtained from the SDSS QSO samples at different redshift bins by using those models listed in Table~\ref{tab:model}, respectively. For each row, left three panels show the results obtained from the same model indicated in the first panel to the left, while the right three panels show those from the same model indicated in the third panel to the right. The region enclosed by the dash yellow (blue) and solid cyan (magenta) lines show the $1\sigma$ and $2\sigma$ confidence range for those cases by adopting the H$\beta$ (or MgII) mass estimator, respectively. The solid green (or red) lines show the fitted value of the likelihood of the correlations for QSOs with H$\beta$ (or MgII) mass estimator.
}
\label{fig:epsm_one}
\end{figure*}

The MCMC fitting results of the SDSS QSOs samples may depend on the details of the modeling. For example, the choices of essential model parameters, forms of BHMF, correction of the host galaxy contamination, and $f_{\rm col}$, may all affect the fitting results. To explore their possible impacts, here we adopt eight different models, with the underlying assumptions of each model listed in Table~\ref{tab:model}.

All models ignore the host galaxy contamination, expect model M12S2S and M12U2S, of which we have assumed that 
the host galaxy correction is given by \citet{Shen11} and \citet{Schulze15}, respectively. 
According to \citet{Shen11}, this ratio is given by
\be
\frac{L_{\rm host}}{L_{\rm QSO}}=0.8052-1.5502x+0.9121x^2-0.1577x^3,
\label{eq:shen11_bw}
\ee
where $x=\log (L_\nu/{\rm erg~s}^{-1})-44$, and $x+44<45.03$;  while according to \citet[][see their Fig. 2]{Schulze15}, it is given by
\be
\frac{L_{\rm host}}{L_{\rm QSO}}=0.7917-0.3405x+0.05689x^2-0.01461x^3,
\label{eq:schulze15_bw}
\ee
where $x+44<46.67$.

The above two fitting formulas are valid only when $\log$ $(L_\nu/{\rm erg~s}^{-1})\ga44$ \citep{Shen11} and $\log$ $(L_\nu/{\rm erg~s}^{-1})\ga 43.8$, respectively. By adopting these corrections, we can recover the intrinsic QSO luminosity, 
and use the MCMC model to extract the intrinsic correlations from these samples.

\subsection{The intrinsic $\epsilon-\bh$ relation}
\label{subsec:sdss_int_cor}

\begin{figure*}
\includegraphics[scale=1.2]{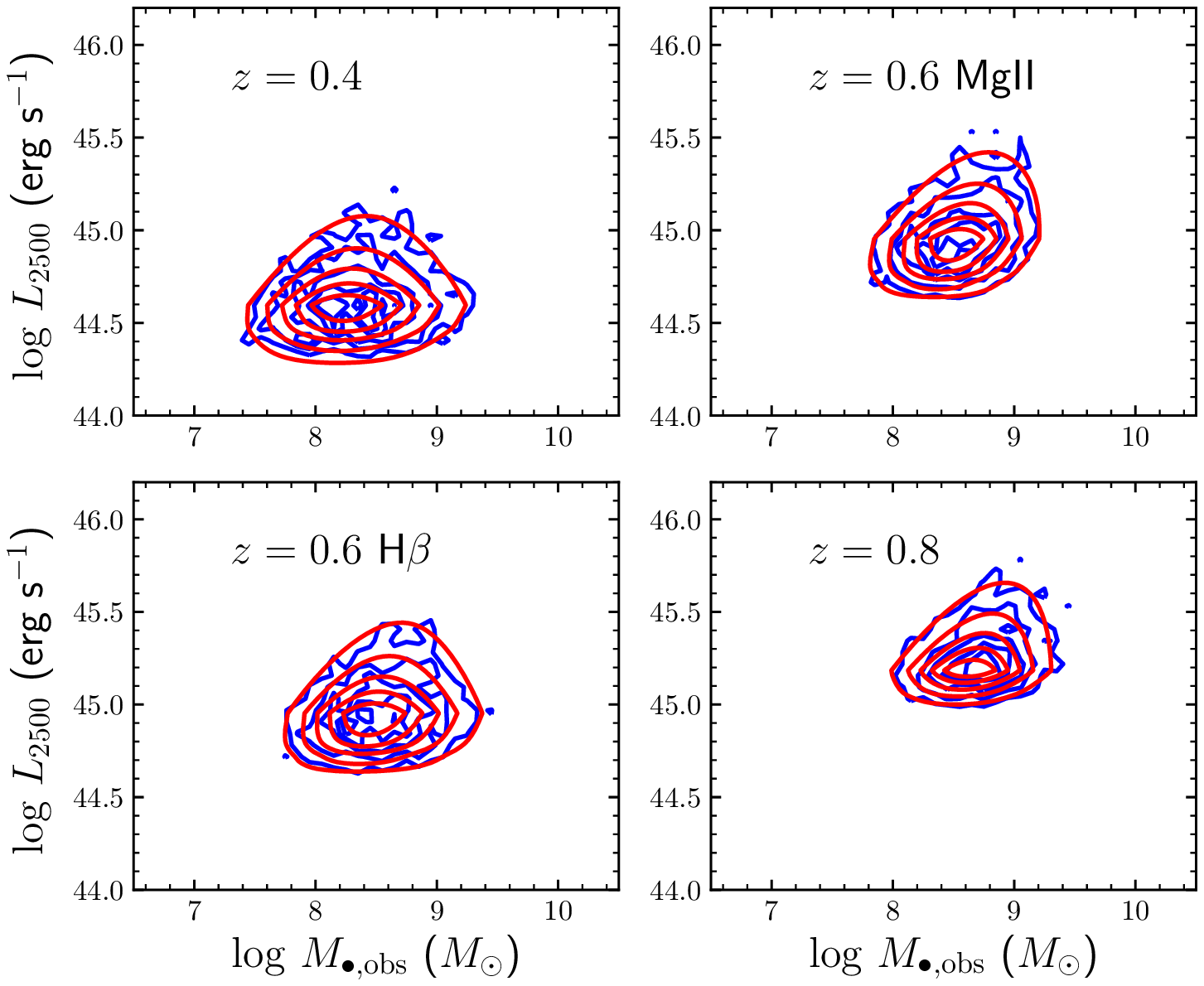}
\caption{
Distributions of QSOs at different redshift bins ($z=0.4$, $0.6$, and $0.8$) on the $L_{2500\AA}-M_{\bullet,\rm obs}$ plane. In each panel, blue and red contours show the observational distributions of SDSS QSOs and the best-fit results
 obtained from the maximum likelihood model of M12N2S (as examples), respectively. {%\color{magenta}
 	For other models in Table 1, the observational distributions can also be well fit}. Note that in redshift bin $z=0.6$, we show the results obtained by using MBH masses estimated from the MgII and H$\beta$ estimators,
respectively. {
	The contour levels from inside to outside represent those with relative number density per pixel values of $5/6, 2/3, 1/2, 1/3,$ and  $1/6$ of peak value, respectively.}
}
\label{fig:mobs_lopt}
\end{figure*}

The intrinsic $\epsilon-\bh$ relation is assumed to be a power-law with slope $\alpha_\epsilon$ and  normalization $c_\epsilon$ at $M_{\bullet}=10^8M_{\odot}$ (see Eq.~\ref{eq:epsilonM}). The best fits of these two parameters obtained from different models are summarized in Table~\ref{tab:result} and shown also in Figure~\ref{fig:sdss_epsilon}. Since the constraints on $c_\epsilon$ is generally weak, it is more clear to see the confidence level of the intrinsic relation in the $\epsilon-\bh$ plane, which is shown in Figure~\ref{fig:epsm_one}.

For the QSO samples utilizing the H$\beta$ mass estimator, we summarize our main results at $0.34<z<0.7$ as following. As shown in Table~\ref{tab:result}, the most likely value of $\alpha_\epsilon$ is $ 0.3\sim1.1$ obtained from the QSO sample with redshift $0.34<z<0.5$ by using different models, and it is still positive but slightly smaller ($0.0\sim0.8$) from the QSO sample with $0.5<z<0.7$. The most likely values of $c_\epsilon$ are $-1.4 \sim -0.7$ obtained from the QSO samples with $0.34<z<0.7$. Apparently, most models suggest a positive correlation between $\epsilon$ and $\bh$, i.e., the larger the MBH mass, the higher $\epsilon$ (or spin). As also seen from Figure~\ref{fig:epsm_one}, the constraint on the $\epsilon-\bh$ relation is quite tight when it is obtained from the QSO sample with $0.34<z<0.5$ (panels in the first and fourth columns), while it has larger uncertainty when obtained from the QSO sample with $0.5<z<0.7$. For all models, $\epsilon$ is low at $\bh<10^7\msun-10^8\msun$ and becomes high  with $\epsilon\sim0.4$ at $\bh>10^9\msun-10^{10}\msun$.

For the QSO samples utilizing the MgII mass estimator, we summarize the main results at $0.5<z<0.7$ as following. The best-fit values of $\alpha_\epsilon = -1.0\sim -0.0$ and $c_\epsilon =-1.0\sim-0.6$. Most of those models tend to result in a negative $\epsilon-\bh$ correlation, though only four among the eight models have $\alpha_\epsilon>0$ at $2\sigma$ confidence level.
The constraints on the $\epsilon-\bh$ relation shown in Figure~\ref{fig:epsm_one} clearly suggest that $\epsilon$ is large when $\bh\sim10^7\msun-10^8\msun$ and small $\epsilon$ when $\bh\sim10^9-10^{10}\msun$. For most models, the obtained results are inconsistent (at $2\sigma$ level) with those constraints obtained from the QSO sample in the same redshift bin but utilizing the H$\beta$ mass estimator.
Note that models like M13N2S, M11N2SG or M13N2SG have very weak constraints on the correlations, thus it seems that they can still be consistent with those utilizing the H$\beta$ mass estimator.

As seen from Figure~\ref{fig:epsm_one} (the third and sixth column panels), almost all models for the QSO sample with $0.7<z<0.9$ result in a constraint  on  $\epsilon-\bh$ relation with which the $2\sigma$ contour covers almost the entire parameter space. Thus, $\alpha_\epsilon$ and $c_\epsilon$ cannot be well constrained if $z\ga 0.7$, as expected and predicted in Section~\ref{subsec:flux_limit} for SDSS QSOs at such redshift range or higher redshifts.

Figure~\ref{fig:mobs_lopt} shows the comparison between the best-fits obtained from model M12S2S (as examples) and observational distributions of QSOs in the $L_{2500} - M_{\bullet,\rm obs}$ plane. We can see that they are well consistent with each other (The observational distributions can also be well fit by other models in Table 1). The $L_\nu-\bh$ joint distribution obtained by using 
the MgII estimator is clearly narrower in mass direction (top right panel of Fig.~\ref{fig:mobs_lopt}) than that using the
H$\beta$ estimator (bottom left panel of Fig.~\ref{fig:mobs_lopt}). Such a difference is similar to that between the cases with $\alpha=-1$ and $\alpha=1$ shown in Figure~\ref{fig:ill_mobs_lopt}. This explains why we always get the most likelihood value of positive $\alpha_\epsilon$ for the QSO sample using the H$\beta$ estimator but a negative $\alpha_\epsilon$ for those samples using the MgII estimator. 

\subsection{The bias due to the mass estimators, host galaxy contamination and other parameters in the model} 
~\label{subsec:biases_mass_estimator}
If inconsistencies between the results from H$\beta$ and Mg II mass estimator are due to 
the bias of luminosity, then including $\beta_{\rm L}$ should make their fitting results consistent with each other. 
However, we do not see such consistency within $2\sigma$ confidence level for those models 
including $\beta_L$ (e.g., M13N2S, M13N2SG), but we see larger uncertainties in the recovered $\epsilon-\bh$ 
correlation.
Such inconsistency are also not disappeared by including a constant $b_{M_{\bullet}}$ as a free parameter 
or not in the MCMC simulations, although $b_{M_{\bullet}}$ is usually non-zero ($\pm0.2\sim0.4$, see Fig.~\ref{fig:sdss_other}) resulting from all these models. 

We find that our above conclusions on the $\epsilon-\bh$ relation for SDSS QSOs are not affected much by including host galaxy contamination or not in the model. This is possibly because the complexities of host galaxy contamination functions. For example, the adopted two kinds of  host galaxy contamination may be either overestimated or underestimated. On the other hand, the host contamination may vary from galaxy to galaxy, the simple functional correction shown 
in Equations~\eqref{eq:shen11_bw} and \eqref{eq:schulze15_bw} may be not accurate enough.

We also find that our conclusions on $\epsilon-\bh$ correlation does not affected by changing $f_{\rm col}=1.7$ to
$f_{\rm col}=1$, or adopting a different distribution function of $\Phi(\bh, \lambda)$
(e.g., using Equation~\ref{eq:philambdadp_gauss} rather than Equation~\ref{eq:philambda}) in the modeling. 
For all the eight models listed in Table 1, we have also try other variations of them, e.g., changing the form of 
the Equation~\ref{eq:phim} in model M13N2S to double power law model (See its footnote); or make host galaxy 
correction by Equation~\ref{eq:shen11_bw} in addition to model M13N2S, we do not see significant 
differences in the recovered correlations. These results suggest that the inconsistencies are not due to these 
complexities. Note that only a few cases were test here and the above conclusion may need further confirmation by
using a more sophisticated model, and etc, in the future.

Thus, the inconsistency between the constraints on $\epsilon-\bh$ relation obtained by using the H$\beta$ estimator and 
that using the MgII one may be partly due to the possible bias induced by using FWHM in the same way for H$\beta$ and 
MgII in the estimator. It is commonly assumed that the MgII and H$\beta$ emitting regions are the same and have the 
same line-width \citep[e.g.,][]{Onken08,Salviander07}. In the SDSSs sample we adopted, they use a fix ratio of $2$ 
for FWHM in both the H$\beta$ and MgII estimators \citep{Shen11}. However, observations suggest that it can be 
different from $2$, as they may be from different line-emitting locations in the BLR \citep[e.g.,][]{Wang09}. 
Currently, it is still unclear whether the estimates of $\bh$ base on MgII consistent with those of based on 
H$\beta$. In the current work, we cannot include FWHM in our modeling, thus, if there is some systematics in 
the FWHM, we cannot reduce them here.

\subsection{Constraints of BHMF and ERDF}
\label{subsec:BHMF_ERDF}

\begin{figure*}
\includegraphics[scale=0.6]{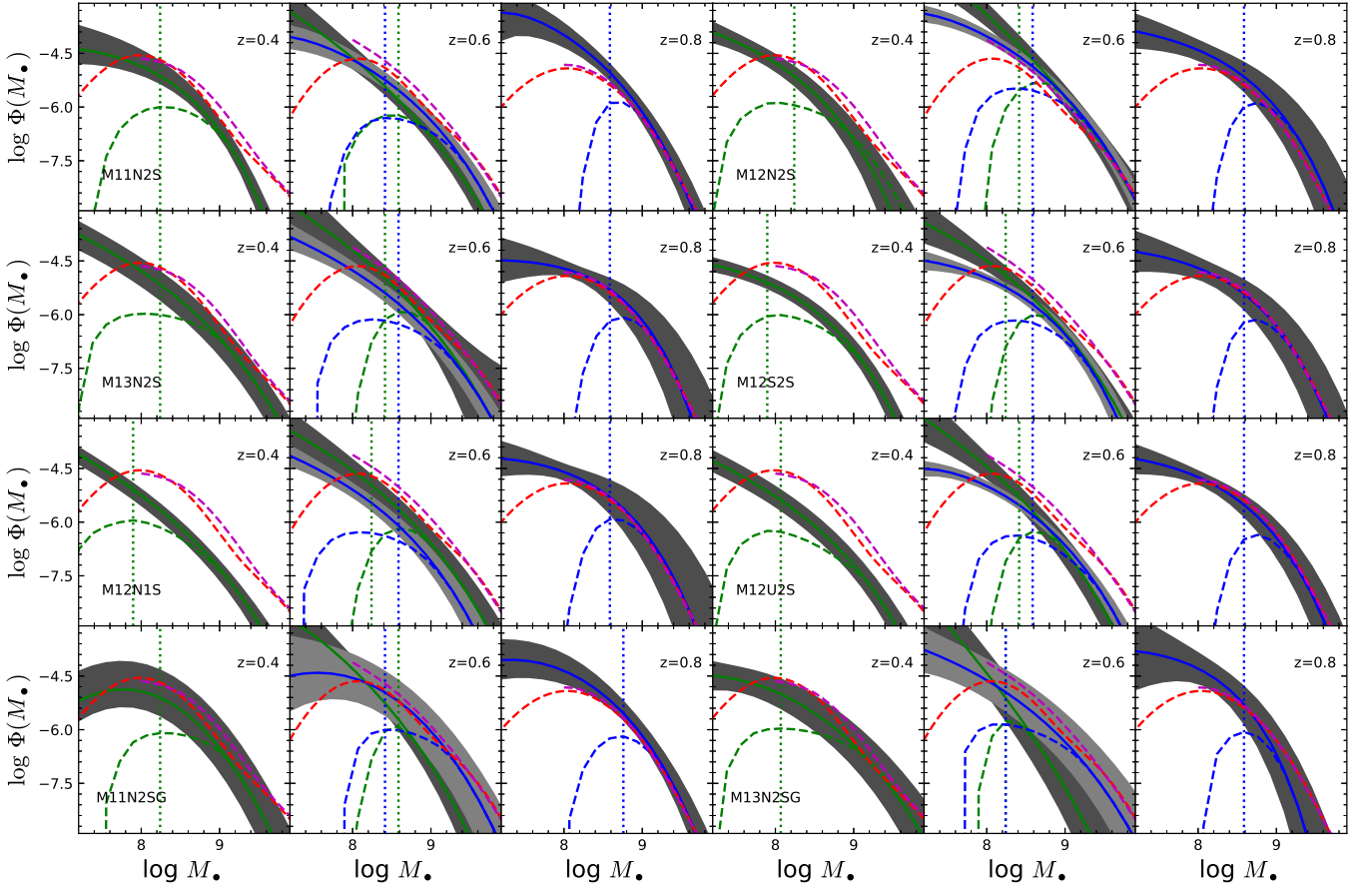}
\caption{
The intrinsic BHMFs. $\log$ $\Phi(\bh)$. Solid lines represent the best-fits and the shaded regions around it show the $1\sigma$ confidence level.
%Green and/or blue solid/dash lines show the intrinsic/observed BHMFs, respectively. 
	Blue and green lines show the results obtained from QSOs samples by utilizing the MgII and H$\beta$ mass estimators, respectively, and the green/blue solid lines show the intrinsic BHMFs obtained from the MCMC fittings, while the green/blue dash lines show the observed BHMFs.
Red and magenta dash lines in each panel 
show the results obtained in \citet{Shen12} and \citet{Kelly13}, respectively. The vertical green and blue dot
lines mark the boundaries below which the completeness of observations is smaller than $10\%$ for QSO samples utilizing the H$\beta$ and Mg II mass estimators, respectively, due to the flux limit.
}
\label{fig:phi_m}
\end{figure*}

\begin{figure*}
\includegraphics[scale=0.6]{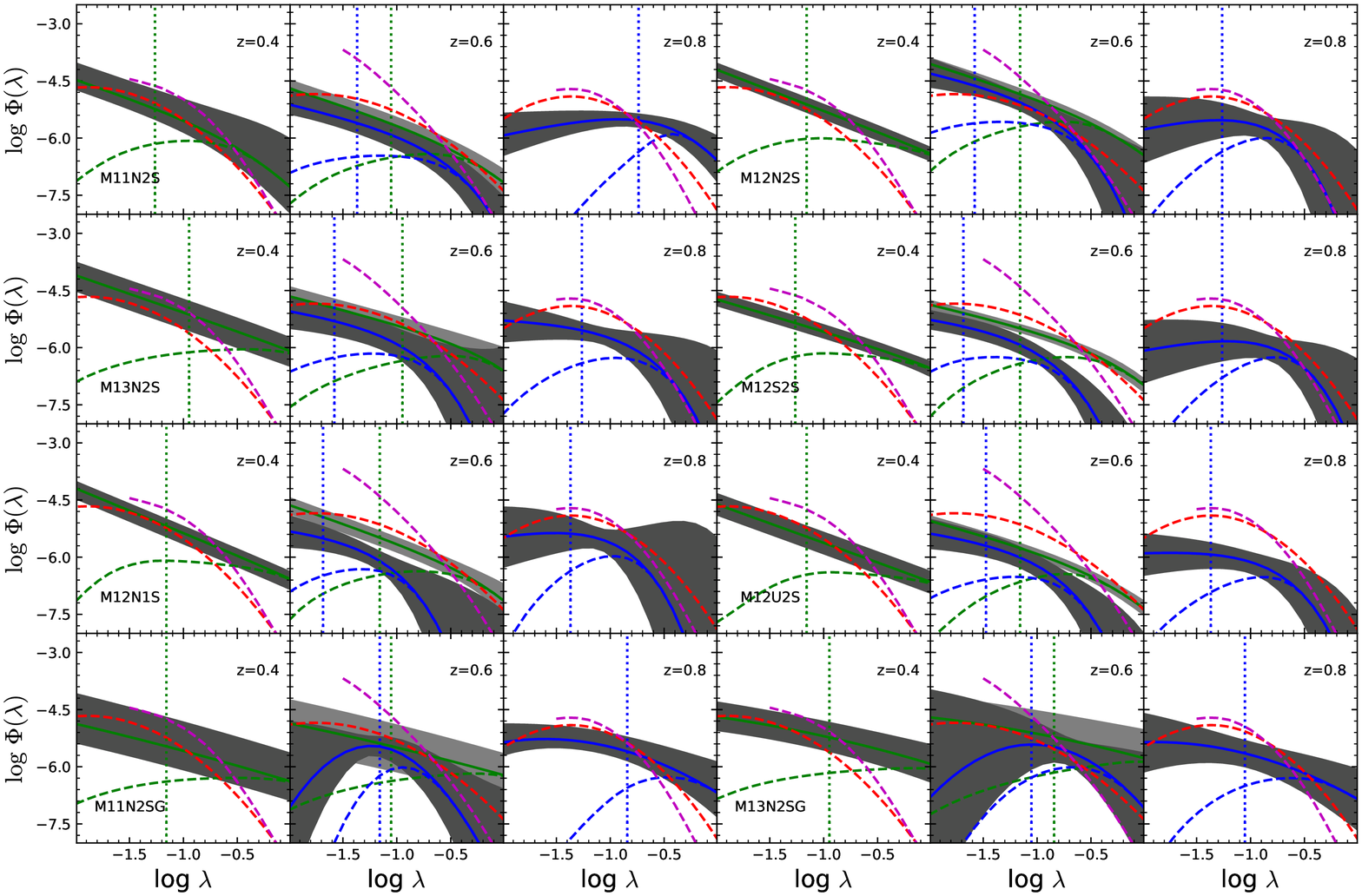}
\caption{Similar to Fig.~\ref{fig:phi_m} but for Eddington ratio.
}
\label{fig:phi_lambda}
\end{figure*}

In addition to the constraint on the $\epsilon-\bh$ relation, even strong constraints on the BHMF and ERDF at redshift $0.34<z<0.9$ can be also obtained from our model. The marginalized intrinsic BHMF and ERDF of the QSOs in each redshift bin are given by
\be
\ba
\Phi(\bh)&= \frac{
\int  \Phi (\bh, \lambda) d\lambda \int \Phi_z(z) \frac{dV}{dz}  dz}
{\int  \frac{dV}{dz} dz},
\ea
\ee
and
\be
\ba
\Phi(\lambda)&=\frac{\int \Phi (\bh, \lambda) d\bh \int \Phi_z(z) \frac{dV}{dz}  dz}
{\int  \frac{dV}{dz} dz},
\label{eq:fphilambda}
\ea
\ee
respectively.

Due to the flux limit, only those QSOs above the luminosity limits can be observed. The observed marginalized intrinsic BHMF and ERDF of the QSOs in each redshift bin is
given by
\be
\ba
\Phi_{\rm obs}(\bh)&= \frac{
\int \Phi (\bh, L_\nu)\Omega(L_\nu,z) dL_\nu \int \Phi_z(z) \frac{dV}{dz} dz}
{\int  \frac{dV}{dz} dz},
\ea
\ee
and
\be
\ba
\Phi_{\rm obs}(\lambda)&=\frac{\int \Phi (\bh, \lambda)\Omega(L_\nu,z) d\bh\int \Phi_z(z) \frac{dV}{dz} dz} {\int  \frac{dV}{dz} dz},
\label{eq:phiobslambda}
\ea
\ee
respectively. Note that the results of $\Phi(\lambda)$ and $\Phi_{\rm  obs}(\lambda)$ depend strongly on the lower limit of the integration of $\bh$, especially if $\lambda$ is small.
It may be not so meaningful to consider the ERDF for MBHs with mass substantially below $\log(\bh/\msun)\sim7-8$, as the completeness of $\bh$ drops rapidly below $10^8\msun$ (see Fig.~\ref{fig:phi_m}).
To exclude the parameter ranges where the QSO samples are highly incomplete, we set the lower limit of $\bh$ for the integrations in above Equations~\eqref{eq:fphilambda} and \eqref{eq:phiobslambda}, below which the completeness of $\bh$, i.e., $\Phi_{\rm obs}(\bh)/\Phi(\bh)$, $\lesssim 10^{-3}$. This lower limit is $\log (\bh/\msun)\sim 7-7.5$ when $z\sim0.4$, $\sim 7.5-8$ when  $z\sim0.6$, and $\sim 8-8.5$ in redshift $z\sim0.8$.

Figure~\ref{fig:phi_m} shows the BHMFs at different redshift obtained from all the eight models listed in Table~\ref{tab:model}. As seen from this Figure, the resulting BHMFs are slightly model dependent for QSOs at $z<0.7$, and some of them are significantly different from  those given by \citet{Shen12} and \citet{Kelly13}. This may suggest that, including the $\epsilon-\bh$ relation and the 
	systematic errors in the estimated MBH masses in the model can affect the determination of BHMFs (similarly
the ERDFs, see below). We find that, the larger the slope $\alpha_\epsilon$ of the $\epsilon-\bh$ relation, the 
larger the difference between the resulting BHMF and that given by \citet{Shen12} and \citet{Kelly13}. 
For example, the models M11N2SG and M13N2SG result in small value of $\alpha_\epsilon$,
and the resulting BHMFs are very close to those given by \citet{Shen12} and \citet{Kelly13}. For 
other models, the difference are more apparent as the $\alpha_\epsilon$ of them is larger.

Figure~\ref{fig:phi_lambda} shows the ERDFs in different redshift bins obtained from our models. As seen from this Figure, the resulting ERDFs are also slightly model dependent. In the redshift bin $0.3<z<0.5$, the ERDFs obtained from our models have a relatively shallower slope than and do not drop rapidly around $\lambda=1$ as those in \citet{Shen12} and \citet{Kelly13}. This is partly because we set a boundary for the Eddington ratio at $\lambda=1$, and partly because of the positive $\epsilon-\bh$ correlations. Similar results for ERDFs are also obtained from the QSO sample with $0.5<z<0.7$ utilizing the H$\beta$ mass estimator. For the QSO sample with $0.5<z<0.7$ utilizing the MgII mass estimator, the resulting ERDFs drop rapidly near $\lambda=1$, and  negative (rather than positive) $\epsilon-\bh$ correlations are obtained from different models. These results again suggest that a none-zero $\epsilon-\bh$ correlation can affect the constraints on ERDF as well as BHMF and raise the importance of considering the $\epsilon-\bh$ relation.
%}

In the redshift bin $0.5<z<0.7$, both the BHMFs and ERDFs resulting from the fittings by utilizing the H$\beta$ (lines in green) and MgII (lines in blue) mass estimators have some discrepancies. Such discrepancies are likely due to the fact that the 
H$\beta$ and MgII mass estimators are not well modeled. As all parameters describing the BHMF, ERDF, and the 
$\epsilon-\bh$ relation are set free in the MCMC fittings, a biased modeling of the mass estimator will then 
result in biased $\epsilon-\bh$ relation (as shown in Section~\ref{subsec:biases_mass_estimator}), BHMF and ERDF. 
In the future, we expect that such discrepancies may disappear if we have more better modeling 
of the mass estimators. 

Note that the way to obtain $\lambda$ in our method is very different from those in \citet{Kelly13} and~\citet{Schulze15}. The Eddington ratio $\lambda$ is derived directly from the optical luminosity and the mass of QSOs in \citet{Kelly13} and \citet{Schulze15}, i.e., $\lambda \propto L_\nu \bh^{-1}$ as they adopt the empirical bolometric correction to directly get the total luminosity. In their methods, the Eddington ratio can be obtained for each sample object. In our method, $\lambda$ of each object cannot be separately obtained from observations and $\epsilon$ of each individual object is unknown. In addition, in our method $\lambda\propto L_\nu^{3/2} \bh^{-2}\mathcal{K}\epsilon$, is also different from those adopted in \citet{Kelly13} and \citet{Schulze15}, according to Equation~\eqref{eq:loptad}.

We notice that \citet{Shen12} and \citet{Kelly13} adopts different modeling between the MBH and luminosity, and their constraints on ERDF is significantly different from each other, especially at high Eddington ratio end. In our model, the Eddington ratio is generally complete if $ \lambda>-1.5\sim-1.0$ for the QSOs at redshift $0.34<z<0.9$. Note that this completeness is for only the samples that the completeness of the black hole masses are also larger than $10^{-3}$.

\subsection{Fitting results of other parameters}

\begin{figure*}
\begin{center}
\includegraphics[scale=0.71]{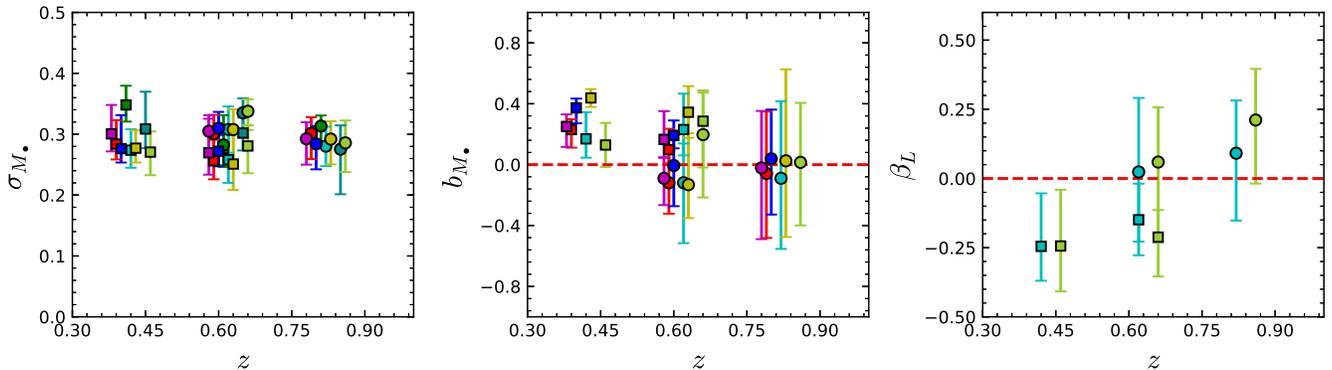}
\caption{
Similar to Figure~\ref{fig:sdss_epsilon} but for model parameters { $\sigma_{M_{\bullet}}$, $b_{M_{\bullet}}$ and $\beta_L$}.
}
\label{fig:sdss_other}
\end{center}
\end{figure*}

The maximum likelihood method applied here can put constraints on not only the intrinsic $\epsilon-\bh$ relation, BHMF, and ERDF, but also simultaneously the other parameters in the model, e.g., such as the parameters related to the mass estimators.
Figure~\ref{fig:sdss_other} shows that our best-fit mass scatter $\sigma_{M_{\bullet}}$ for QSOs with $0.3<z<1.5$ is around $\sim0.3$ dex, consistent with the value $0.3-0.4$ dex that commonly used in the literature~\citep[e.g.,][]{Schulze15,Kelly10}. 

According to those models with $13$-parameters, we find that the constant mass bias $b_{M_{\bullet}}$ and 
the possible luminosity-dependent systematic errors introduced by $\beta_{\rm L}$ in the H$\beta$ and MgII estimators may be different. For the QSO sample with $0.34<z<0.7$, 
all models result in $b_{M_{\bullet}} \sim 0.1-0.4$, $\beta_{\rm L}=-0.25\sim-0.15$ by using the H$\beta$ estimator, 
while for the QSO sample with $0.5<z<0.7$, they result in $b_{M_{\bullet}} \sim -0.1-0.2$, 
$\beta_{\rm L}\sim0.02-0.2$ by using the MgII estimator. 
These results suggest that both the constant mass bias and the luminosity-dependent systematic error 
in the MgII estimator may be different with those in the H$\beta$ one.

\section{discussions}
\label{sec:dis}
%For a comprehensive picture, it is quite important for us to tie the other parameter (other than those
%in the mass-epsilon relation and the mass estimators) and see if the correlations are consistent
%using the two mass estimators. Such test can only be done by a more comprehensive modeling, 
%that includes both the observed masses of samples using H$\beta$ and MgII 
%in the likelihood function (Eq 9). However, we think that the accuracy of such framework depends largely 
%on the correctness of the modeling of individual mass estimators. 
%For example, if the systematics of one (or both) of the mass estimators is not accurately modeled, 
%then the constraints from MCMC by fitting to both of them will be problematic. 
%Thus, it will be quite important for us to first understand correctly the systematic of individual mass 
%estimators before we can take further steps to combine both of them. 
%Such an extension of the current work will merits a whole new paper to calculate and discuss about it. 
%Considering such complexity, we do not intented to do it in the current work.

Due to the discrepancies between results obtained by using the Mg II and H$\beta$ mass estimators,
it is difficult to put solid conclusions on the $\epsilon-\bh$ relation, the BHMF and ERDF of the SDSS QSOs. 
If both of the two mass esimators are well-modeled, we expect that these results should be consistent with each other.
If not, both the $\epsilon-\bh$ relation and the BHMF or ERDF would depend on the correctness of mass estimators, 
and they will be biased if there are any remaining systematics. Such unknown systematics could possibly be addressed 
in the future, by extending our models to include FWHM, velocity dispersion, or other details of the mass estimator(s)
[similar to some previous works, e.g.,][]\citep{Kelly09}. On the other hand, it should be straightforward 
to expending our model likelihood function (Eq.~\ref{eq:maxim_likelihood}) to fit QSOs samples with masses from 
two different mass estimators with equal weight. 
%Such that we can fit simultaneously to QSOs with masses obtained from both the MgII and H$\beta$ mass estimators.
However, the accuracy of such a model is unknown, as currently we do not known whether the $\epsilon-\bh$
relations obtained by using the MgII and H$\beta$ mass estimators are biased to negative or positive values. Only in the cases
that the MgII and H$\beta$ mass estimators are biased in opposite ways we can get more accurate results. 
Thus, this problem may be solved if the uncertainties in the Mg II and Hbeta mass estimators can be well 
determined or a better mass estimator can be provided, in which case more robust and consistent constraints on the 
$\epsilon-\bh$ relation can be obtained.

The $\epsilon-\bh$ relation of QSOs has also been investigated and discussed in some preivous works but in a way different from the maximum likelihood method introduced in this paper. In these works, the radiation efficiency $\epsilon$ of individual QSOs were mostly estimated according to the QSO bolometric luminosities and the inferred accrection rates \citep[e.g.,][]{Davis11, Wu13, Raimundo12, Trakhtenbrot17, Schulze17, Shankar19}. However, the selection effects of the flux limits of QSO surveys must be carefully considered, as it may lead to an apparent correlation between $\epsilon$ and $\bh$, even if there is no intrinsic one \citep[e.g., see][]{Wu13, Raimundo12}. Furthermore, the model adopted in these works is not physical and self-consistent, as the averaged bolometric correction is used for those QSOs with the same luminosity at a given band but different masses and accretion rates. According to the accretion disk model, the bolometric corrections for QSOs should be dependent on the masses, spin, and accretion rate
\citep[see more discussions in][]{Netzer14}. As a comparison, our work is a statistical method, in which we do not focus on individual QSOs but the statistical distributions of QSO samples on the mass-luminosity plane. In our method, we have also taken account of the complexities due to the selection functions and the biases in the mass estimators, which is self-consistent as demonstrated by using the mock data.

The $\epsilon-\bh$ correlation may be also inferred from the (cosmological) evolution of black hole spins and mean radiative efficiencies \citep[e.g.,][]{Li5, ZL19}. The cosmological growth of the QSO masses and total luminosity radiated from QSOs are directly linked by the (mean) radiative efficiency, and thus the radiative efficiency (or spin) can be constrained according to the cosmological evolution of black hole mass density and the total radiated energy inferred from the QSO luminosity functions [see the So{\l}tan argument in \citet{Soltan82} and the extended one in \citet{YL04}; the possible evidence for cosmological evolution and mass dependent of radiative efficiency in \citet{Li5, WangJM09}]. However, the results on the $\epsilon-\bh$ relation possibly obtained by using the above method suffer from the large uncertainties in the empirical relations between black hole mass and galaxy properties that are adopted to estimate black hole mass densities and the uncertainties in the bolometric corrections and the luminosity functions used to estimate the QSO total energy densities. Considering also that our current model does not include the detailed physics to trace the cosmological black hole growth, spin and radiative efficiency evolution, it would be important to cross check future results on the $\epsilon-\bh$ relation obtained from the method presented in the current paper with those from that mentioned above. 

\section{Conclusions}
\label{sec:con}

The intrinsic relation between radiative efficiency ($\epsilon$) and mass of MBHs ($\bh$) in QSOs, if any, is important for understanding the MBH/QSO cosmological evolutions. In this paper, we develop a maximum likelihood method to extract the intrinsic $\epsilon-\bh$ relation from the distribution of QSOs in the optical luminosity-MBH mass plane. We adopt a simple relativistic thin accretion disk model to estimate the radiation of QSOs in the optical-UV band. 

We first use mock QSO samples to demonstrate that the intrinsic relation, if any, can be extracted from a SDSS-DR7-like QSO sample with $z\sim 0.4-0.6$. For QSOs at higher redshift with $0.6\la z\la0.9$, the intrinsic relation can be robustly extracted only if a QSO survey can go about $\sim 1-2$ magnitude ($i_{\rm mag} \simeq20-21$ mag) fainter than the current SDSS survey. If a SDSS-like survey with flux limit down to $\simeq22$\,mag, then it is possible to identify the intrinsic $\epsilon-\bh$ relation for a QSO sample at redshift upto $z\sim 1.5$.

We find that such an intrinsic relation can still be extracted even if it has a scatter. Under some 
circumstances, if the samples are contaminated by host galaxy and it is not corrected, 
the recovered $\epsilon-\bh$ relation may be significantly biased from the intrinsic ones. 

We apply our method to the SDSS QSOs in the redshift range $0.34<z<0.9$ and adopt a number of models to perform MCMC simulations and get constraints on the intrinsic $\epsilon-\bh$ relation. We find signs of positive $\epsilon-\bh$ correlations ($\epsilon \propto M_{\bullet}^{0\sim 1.1}$) for QSOs at $0.34<z<0.7$ utilizing the H$\beta$ mass estimator. However, 
for QSOs at $0.5<z<0.7$ utilizing the Mg II mass estimator, we find that the correlations become negative 
($\epsilon \propto M_{\bullet}^{-1.0\sim 0}$). This inconsistency is likely due to the unknown different systematic errors in these two mass estimators, but not due to the host galaxy contamination. 

We also derive constraints on the BHMF and ERDF of QSOs. 
Although our results are broadly consistent with some previous studies, e.g.,
\citet{Shen12} and \citet{Kelly13}, we find that the $\epsilon-\bh$ relation, if does exist, can affect the determination 
of BHMF and ERDF significantly; 
This may suggest that, to obtain precise constraints on both the BHMF and ERDF, it is necessary to include the $\epsilon-\bh$ relation in the modeling. 

 We conclude here that the maximum likelihood method introduced in this paper can set robust constraints on the intrinsic $\epsilon-\bh$ relation, if any. However, if there are some unknown systematic errors in the mass estimators that cannot be well modeled, the recovered $\epsilon-\bh$ relation may be strongly biased from the true one. We expect that the understanding of the systematics on the MBH mass estimator(s) will be improved a lot and the QSO surveys will go much deeper than the SDSS QSO survey adopted here, which will enable the revealing of the intrinsic $\epsilon-\bh$ relation and better determinations of the BHMF and the ERDF of QSOs.
 In addition, it is possible to solve this problem if we can improve our model to consider the complexities of the FWHM of the mass estimators.

\acknowledgements{
%\noindent
We thank the anonymous referee for the helpful comments that have improved this paper. 
We thank Dou Li-Ming for helpful discussions. This work was supported in part by the National Natural Science Foundation of China under grant No. 11603083, 11733010, 11873056, 11991052, the National Key Program for Science and Technology Research and Development (Grant No. 2016YFA0400704), Guangzhou University Startup Fund. The simulations in this work were performed partly in the TianHe II National Supercomputer Center in Guangzhou, and partly on the computing cluster in School of Physics and Astronomy, Sun Yat-Sen University.
}


\begin{thebibliography}{1}
%
\bibitem[Bardeen et al.(1972)]{Bardeen72} Bardeen, J.~M., Press, W.~H., \&
Teukolsky, S.~A.\ 1972, \apj, 178, 347
%
\bibitem[Bentz et al.(2009)]{Bentz09} Bentz, M.~C., Peterson, B.~M.,
Netzer, H., Pogge, R.~W., \& Vestergaard, M.\ 2009, \apj, 697, 160
%
%\bibitem[Goulding et al.(2014)]{Goulding14} Goulding, A.~D., Forman,
%W.~R., Hickox, R.~C., et al.\ 2014, \apj, 783, 40
%
\bibitem[Cunningham(1975)]{Cunningham75} Cunningham, C.~T.\ 1975, \apj, 202, 788

\bibitem[Dauser et al.(2010)]{Dauser10} Dauser, T., Wilms, J.,
Reynolds, C.~S., \& Brenneman, L.~W.\ 2010, \mnras, 409, 1534
%
\bibitem[Davis \& Laor(2011)]{Davis11} Davis, S.\ W., Laor, A., 2011,
ApJ, 728, 98
%
\bibitem[Dotti et al.(2013)]{Dotti13} Dotti, M., Colpi, M., Pallini,
S., Perego, A., \& Volonteri, M.\ 2013, \apj, 762, 68
%
\bibitem[Dubois et al.(2014)]{Dubois14} Dubois, Y., Volonteri, M., \&
Silk, J.\ 2014, \mnras, 440, 1590
%
\bibitem[Gavignaud et al.(2008)]{Gavignaud08} Gavignaud, I., Wisotzki,
L., Bongiorno, A., et al.\ 2008, \aap, 492, 637
%
\bibitem[Gierli{\'n}ski et al.(2001)]{Gierlinski01} Gierli{\'n}ski,
M., Macio{\l}ek-Nied{\'z}wiecki, A., \& Ebisawa, K.\ 2001, \mnras,
325, 1253
%
\bibitem[Hubeny \& Lanz(1995)]{Hubeny95} Hubeny, I., \& Lanz, T.\
1995, \apj, 439, 875
%
\bibitem[Kelly et al.(2009)]{Kelly09} Kelly, B.~C., Vestergaard, M., \& Fan, X.\ 2009, \apj, 692, 1388

\bibitem[Kelly et al.(2010)]{Kelly10} Kelly, B.~C., Vestergaard, M.,
Fan, X., et al.\ 2010, \apj, 719, 1315
%
\bibitem[Kelly \& Shen(2013)]{Kelly13} Kelly, B.~C., \& Shen, Y.\
2013, \apj, 764, 45
%
\bibitem[King \& Pringle(2006)]{kin06} King, A.~R., \& Pringle, J.~E.\
2006, \mnras, 373, L90
%
\bibitem[Kochanek et al.(2012)]{Kochanek12} Kochanek, C.~S.,
Eisenstein, D.~J., Cool, R.~J., et al.\ 2012, \apjs, 200, 8
%
\bibitem[Kollmeier et al.(2006)]{Kollmeier06} Kollmeier, J.~A., Onken,
C.~A., Kochanek, C.~S., et al.\ 2006, \apj, 648, 128
%
\bibitem[Krolik(1999)]{Krolik99} Krolik, J.\ H., 1999, Active Galactic
Nuclei: from the central black hole to the galactic environment
(Princeton, Princeton University Press)
%
\bibitem[Li et al.(2015)]{Li5} Li, Y.-R., Wang, J.-M., Cheng, C., et al.\ 2015, \apj, 804, 45
%
\bibitem[Lusso et al.(2010)]{Lusso10} Lusso, E., Comastri, A.,
Vignali, C., et al., 2010, A\&A, 2512, 34
%
\bibitem[Marconi et al.(2004)]{Marconi04} Marconi, A., Risaliti, G.,
Gilli, R., et al.\ 2004, \mnras, 351, 169
%
\bibitem[Marshall et al.(1983)]{Marshall83} Marshall, H.~L.,
Tananbaum, H., Avni, Y., \& Zamorani, G.\ 1983, \apj, 269, 35
%
\bibitem[McGill et al.(2008)]{McGill08} McGill, K.~L., Woo, J.-H.,
Treu, T., \& Malkan, M.~A.\ 2008, \apj, 673, 703-714
%
\bibitem[Mclure \& Dunlop (2004)]{Mclure04} McLure, R.~J. \& {Dunlop},
J.~S., 2004, MNRAS, 352,1390
%
\bibitem[Merloni et al.(2010)]{Merloni10} Merloni, A., Bongiorno, A.,
Bolzonella, M., et al.\ 2010, \apj, 708, 137
%
\bibitem[Netzer \& Trakhtenbrot(2014)]{Netzer14} Netzer, H. \& Trakhtenbrot, B.\ 2014, \mnras, 438, 672
%
\bibitem[Novikov \& Thorne(1973)]{NT73} Novikov, I.~D., \& Thorne,
K.~S.\ 1973, Black Holes (Les Astres Occlus), 343
%
\bibitem[Onken \& Kollmeier(2008)]{Onken08} Onken, C.~A., \& Kollmeier,
J.~A.\ 2008, \apjl, 689, L13

\bibitem[Perego et al.(2009)]{per09} Perego, A., Dotti, M., Colpi, M.,
\& Volonteri, M., 2009, \mnras, 399, 2249
%
\bibitem[Raimundo et al.(2012)]{Raimundo12} Raimundo, S.\ I., Fabian,
A.\ C., Vasudevan, R.\ V., Gandhi, P., Wu, J., 2012, MNRAS, 419, 2529
%
\bibitem[Reynolds(2014)]{Reynolds14} Reynolds, C.~S.\ 2014, \ssr, 183,
277
%
\bibitem[Richards et al.(2006)]{Richards06} Richards, G.~T., Strauss,
M.~A., Fan, X., et al.\ 2006, \aj, 131, 2766

\bibitem[Salvesen et al.(2013)]{Salvesen13} Salvesen, G., Miller, J.~M., Reis, R.~C., et al.\ 2013, \mnras, 431, 3510

\bibitem[Salviander et al.(2007)]{Salviander07} Salviander, S., Shields, G.~A., Gebhardt, K., \& Bonning, E.~W.\ 2007, \apj, 662, 131

\bibitem[Shankar et al.(2019)]{Shankar19} Shankar, F., Allevato, V., Bernardi, M., et al.\ 2019, Nature Astronomy, 4, 282

\bibitem[Shen et al.(2011)]{Shen11} Shen, Y., Richards, G.\ T.,
Strauss, M.\ A., et al.\ 2011, ApJS, 194, 45
%
\bibitem[Shen(2013)]{Shen13} Shen, Y.\ 2013, Bulletin of the Astronomical Society of India, 41, 61

\bibitem[Shen \& Kelly(2012)]{Shen12} Shen, Y., \& Kelly, B.~C.\ 2012,
\apj, 746, 169
%
\bibitem[Shimura \& Takahara(1995)]{Shimura95} Shimura, T., \&
Takahara, F.\ 1995, \apj, 445, 780
%
\bibitem[Schulze et al.(2015)]{Schulze15} Schulze, A., Bongiorno, A.,
Gavignaud, I., et al.\ 2015, \mnras, 447, 2085
%
\bibitem[Schulze et al.(2017)]{Schulze17} Schulze, A., Done, C., Lu, Y., et al.\ 2017, \apj, 849, 4
%
\bibitem[So{\l}tan(1982)]{Soltan82} So{\l}tan, A., 1982, MNRAS, 200,
115
%
\bibitem[Shakura \& Sunyaev(1973)]{Shakura73} Shakura, N.\ I.,
Sunyaev, R.\ A., 1973, A\&A, 24, 337
%
\bibitem[Trakhtenbrot et al.(2017)]{Trakhtenbrot17} Trakhtenbrot, B., Volonteri, M., \& Natarajan, P.\ 2017, \apjl, 836, L1
%
\bibitem[Volonteri et al.(2013)]{Volonteri13} Volonteri, M., Sikora,
M., Lasota, J.-P., \& Merloni, A.\ 2013, \apj, 775, 94
%
\bibitem[Wang et al.(2009)]{Wang09} Wang, J.-G., Dong, X.-B.,
Wang, T.-G., et al.\ 2009, \apj, 707, 1334

\bibitem[Wang et al.(2009)]{WangJM09} Wang, J.-M., Hu, C., Li, Y.-R., et al.\ 2009, \apjl, 697, L141

\bibitem[Wu et al.(2013)]{Wu13} Wu, S., Lu, Y., Zhang, F., \& Lu, Y.\
2013, \mnras, 436, 3271
%
\bibitem[Yu \& Lu(2004)]{YL04}Yu, Q., \& Lu, Y.\ 2004, ApJ, 602, 603
%
\bibitem[Yu \& Tremaine(2002)]{YT02} Yu, Q., \& Tremaine, S.\ 2002,
\mnras, 335, 965
%
\bibitem[Zhang et al.(2015)]{Zhang15} Zhang, F., Lu, Y., \& Yu, Q.\
2015, \apj, 809, 127

\bibitem[Zhang, \& Lu(2019)]{ZL19} Zhang, X., \& Lu, Y.\ 2019, \apj, 873, 101

\end{thebibliography}
\end{document}